\documentclass[twocolumn,           
               showpacs,            
               nopreprintnumbers,     
               aps,                 
               prl,          	    
               letterpaper,             
              groupeaddress,      
               nofootinbib,         
               tightenlines,        
               floats,floatfix,      
               showkeys
               ]{revtex4-1}

\usepackage{graphicx}
\usepackage{dcolumn}
\usepackage{bm}
\usepackage{amsmath}
\usepackage{amsfonts,amssymb}
\usepackage{color}
\definecolor{v}{rgb}{0.6, 0.2, 0.8} 
\definecolor{a}{rgb}{0.59, 0.29, 0.} 

\usepackage{enumerate}
\usepackage{float}
\usepackage{subfig}
\usepackage{ulem}
\usepackage{soul}

\begin{document}

\title{Brane with variable tension as a possible solution to the problem of the late cosmic acceleration}

\author{Miguel A. Garc\'{\i}a-Aspeitia$^{1,2}$}
\email{aspeitia@fisica.uaz.edu.mx}

\author{A. Hernandez-Almada$^3$}

\author{Juan Maga\~na$^4$}

\author{Mario H. Amante$^{1,4}$}

\author{V. Motta$^4$}

\author{C. Mart\'inez-Robles$^1$}

\affiliation{$^1$Unidad Acad\'emica de F\'isica, Universidad Aut\'onoma de Zacatecas, Calzada Solidaridad esquina con Paseo a la Bufa S/N C.P. 98060, Zacatecas, M\'exico.}
\affiliation{$^2$Consejo Nacional de Ciencia y Tecnolog\'ia, \\ Av. Insurgentes Sur 1582. Colonia Cr\'edito Constructor, Del. Benito Ju\'arez C.P. 03940, Ciudad de M\'exico, M\'exico.}
\affiliation{$^3$Facultad de Ingenier\'ia, Universidad Aut\'onoma de Quer\'etaro, Centro Universitario Cerro de las Campanas, 76010, Santiago de Quer\'etaro, M\'exico}
\affiliation{$^4$Instituto de F\'isica y Astronom\'ia, Facultad de Ciencias, Universidad de Valpara\'iso, Avda. Gran Breta\~na 1111, Valpara\'iso, Chile.}

\begin{abstract}
Braneworld models have been proposed 
as a possible solution to the problem of the accelerated expansion of the Universe. 
The idea is to dispense the dark energy (DE) and drive the late-time cosmic acceleration with a five-dimensional geometry.
Here, we investigate a brane model with variable brane tension as a function of redshift called chrono-brane. We propose the polynomial $\lambda=(1+z)^{n}$ function
inspired in tracker-scalar-field potentials. To constrain the $n$ exponent we use the latest
observational Hubble data from cosmic chronometers, Type Ia Supernovae from the full JLA sample,
baryon acoustic oscillations and the posterior distance from the cosmic microwave background of Planck 2015 measurements.
A joint analysis of these data estimates $n\simeq6.19$ which  generates a DE-like or cosmological-constant-like term, in the Friedmann equation arising from the  extra dimensions. This model is consistent
with these data and can drive the Universe to an accelerated phase at late times.
\end{abstract}

\date{\today}
\maketitle

\textit{Introduction.-} The accelerated expansion of the Universe in the present epoch is supported by high-resolution observations of Supernovae Type Ia (SNIa) at high redshift \cite{Schmidt,Perlmutter,Riess}, 
anisotropies in cosmic microwave background radiation (CMB) \cite{Planck:2015XIII,Planck:2015XIV} and baryon acoustic oscillations (BAO) \cite{Alam:2017}. To explain this within the General Relativity framework, a negative-pressure fluid, dubbed dark energy (DE), must be postulated to produce the observed gravity repulsion \cite{Copeland:2006wr}.
The most economic attempt comes from the cosmological constant \citep[CC;][]{PlanckCollaboration2013},  originated by quantum vacuum fluctuations, with a theoretical value differing $\sim 120$ orders in magnitude from the cosmological observations \cite{Weinberg,Zeldovich}. 

Extra-dimensions scenarios have been proposed to solve the CC problems such as Brane-world models, which accelerate the Universe under the assumption of a $4+1$-dimensional space-time (the bulk) containing an ordinary $3+1$-dimensional manifold (the brane) through a threshold radius $r_s$. However, the majority of them (including the Randall and Sundrum (RS)\footnote{RS models are divided in the case of two (RSI) and one brane (RSII) respectively.} \cite{Randall-I,Randall-II}
models) achieve a stable late cosmic acceleration only by including DE \cite{Dvali:2000hr,Garcia-Aspeitia:2016kak}. Although models with variable brane tension (VBT), $\lambda(t)\propto a(t)$, sourcing from thermodynamics assumptions (E\"{o}tv\"{o}s law) have been studied previously\footnote{Notice that only models with constant brane tension have been observational constrained.} \cite{HoffdaSilva:2011bd,Guendelman:2002mf,Gergely:2008jr,Aros:2016wpv,Bazeia:2014tua,Casadio:2013uma,daRocha:2012pt,PhysRevD.80.046003,Casadio:2016P},
they either have not been contrasted with recent observations or still need the introduction of a DE fluid to reproduce the late acceleration.

In this letter, we propose a phenomenological Brane-world model based on RSII using one brane with variable tension, called \textit{chrono-brane} hereafter, which does not only supply the late-time cosmic acceleration but it is also in agreement with observational data. 

In contrast with previous studies, a double Bianchi identity is not applied, i.e. there is no matter creation into the brane and the modifications appear only at the bulk level. 
Hence, we investigate the effects of a VBT 
in terms of the scale factor (redshift), i.e. $\lambda(a)$ or $\lambda(z)$, on a background cosmology.
We propose a polynomial function for the brane tension which is dominant in later times in the Universe evolution, 
but subdominant in the early Universe to be consistent with the Nucleosynthesis observations.
To probe this cosmological model, we perform a Monte Carlo Markov Chain (MCMC) analysis through the observations of 
SNIa, $\rm H(z)$, BAO, and CMB. We also investigate the scale factor dynamics and the 
cosmological evolution of the different components of the Universe. 

\textit{Braneworld cosmological framework.-} 
The characteristic brane parameter is encoded in the brane tension which establishes the limits where the traditional Einstein's equations are recovered and the terms that comes from extra dimensions become important.
Following Refs. \cite{Gergely:2008fw,Wong:2010rg} the Einstein's field equations projected onto the four dimensional manifold 
with a VBT (see \cite{sms,m2000} for a constant brane tension) can be written in the form:
\begin{equation}
G_{\mu\nu}+\xi_{\mu\nu}=\kappa^2_{(4)}T_{\mu\nu}+\frac{6\kappa^2_{(4)}}{\lambda}\Pi_{\mu\nu}+\sqrt{\frac{6}{\lambda}}\kappa_{(4)}F_{\mu\nu},
\end{equation}
where $\xi_{\mu\nu}$ is the nonlocal Weyl tensor, $T_{\mu\nu}$ and $\Pi_{\mu\nu}$ are the energy-momentum (EM) tensor and the quadratic EM tensor respectively, $\kappa_{(4)}$ is the four dimensional coupling constant and $\lambda$ is the brane tension, being now a function of time (through the scale factor $a$). We also assume no matter fields in the bulk i.e. $F_{\mu\nu}=0$ and no pull back to the brane, 
which is related to the non-standard model fields \cite{Wong:2010rg}. In addition, we consider that the bulk black hole mass vanishes, 
reducing the geometry to AdS$_5$ \cite{m2000}. Therefore, the Friedmann equation can be written as
$\label{eq:H2}H^2=\frac{\kappa^2_{(4)}}{3}\sum_i\rho_i\left(1+\frac{\rho_i}{2\lambda}\right)$,
where $H\equiv\dot{a}/a$ is the Hubble parameter. The deceleration parameter is given by $q(z)=((z+1)/E(z))(dE(z)/dz)-1$. Notice that
the traditional Friedmann equation is recovered when $\lambda\to\infty$. The brane tension can be rewritten as $\lambda(a)\equiv\lambda_0\hat{\lambda}(a)$, with $\lambda_0$ as a free parameter with units [eV]$^4$. The general dimensionless function, $\bar{\lambda}(a)$, gives the brane tension behavior in terms of the scale factor. This way of writing $\lambda$ avoids problems with the fundamental constants (see \cite{Gergely:2008jr} for E\"{o}tv\"{o}s branes), eliminating the temporary dependence, for instance, $\kappa^2_{(4)}=8\pi G_N=\kappa^4_{(5)}\lambda_0/6$, where $\lambda(a)$ has been absorbed by any of the tensors associated with the energy-momentum or with the Weyl's tensor.

The main goal is to source the late cosmic acceleration with a VBT without demanding the presence of DE.
Thus, we only consider matter (baryonic and dark matter) and radiation as the components of the Universe.
A dimensionless Friedmann equation, $E(z)^2 \equiv H^{2}(z)/H^{2} _{0}$, can be re written in terms of the redshift as
\begin{equation}
E(z)^2 = E^{2}_{nb}(z)+
\frac{\mathcal{M}}{\hat{\lambda}(z)}\left[\Omega _{m0}^2  (z+1)^{6} + \Omega _{r0}^2 (z+1)^{8} \right], \label{EI}
\end{equation}
where $E^{2}_{nb}(z)=\Omega _{m0} (z+1)^{3}+ \Omega _{r0} (z+1)^{4}$, $\mathcal{M}=3H_{0}^{2}/2\kappa_{4}^{2}\lambda_{0}$, and $\Omega_{m0}=\Omega_{b0}+\Omega_{DM0}$. 
The radiation component can be expressed as $\Omega_{r0}=2.469\times10^{-5}h^{-2}(1+0.2271N_{eff})$, 
$h=H_{0}/100\mathrm{km}\mathrm{s^{-1}Mpc^{-1}}$, and $N_{eff}=3.04$ is the standard number of relativistic species\footnote{Here we consider $\Omega_{i}=\rho_{i}/\rho_{c}$ and $\rho_{c}$ as the standard critical density.}. From the flatness condition we obtain 
\begin{equation}
\mathcal{M}=\frac{1-\Omega_{m0}-\Omega_{r0}}{\Omega_{m0}^2+\Omega_{r0}^2} \hat{\lambda}(0). \label{eq:Momegas}
\end{equation}
Moreover, the deceleration parameter can be written in the form $q(z) = (q_{I}(z) + q_{II}(z))/E^{2}(z)$, where we define:
\begin{equation}
q_{I}(z) \equiv  \frac{1}{2}\Omega_{m0}(z+1)^3 +\Omega_{r0}(z+1)^4, \label{q1}
\end{equation}
and
\begin{eqnarray}
&&q_{II}(z) \equiv \frac{ \mathcal{M}}{\hat{\lambda}(z)}[2\Omega_{m0}^2(z+1)^6 +3\Omega_{r0}^2(z+1)^8\nonumber\\&&-\frac{1}{2\hat{\lambda}(z)}\frac{d\hat{\lambda}(z)}{dz}[\Omega_{m0}^2(z+1)^7+\Omega_{r0}^2(z+1)^9]].
\label{eq:qII}
\end{eqnarray}
We notice that Eqs. \eqref{EI} and $q(z)$ are reduced to those shown in \cite{Garcia-Aspeitia:2016kak} 
when the brane tension is constant, i.e. $\lambda(z)=\lambda_0$, and when the DE is added.

In order to explore the background cosmology, we propose the following ansatz for the VBT: $\hat{\lambda}(a)=a^{-n}\rightarrow \hat{\lambda}(z)=(z+1)^n$,
where  $n$ $\in \mathbb{R}$ is the free parameter and 
$\hat{\lambda}(1)=\hat{\lambda}(0)=1$ for the scale factor and redshift respectively. Other authors have already analyzed the case $\hat{\lambda}(a)=1-a^{-1}\rightarrow \vert\hat{\lambda}(z)\vert=z$ \cite{HoffdaSilva:2011bd,Guendelman:2002mf,Gergely:2008jr,Aros:2016wpv,Bazeia:2014tua,Casadio:2013uma,daRocha:2012pt,PhysRevD.80.046003,Casadio:2016P}. This form ($n=1$) is inferred through the E\"{o}tv\"{o}s law $\lambda=K(T_c-T)$, where $K$ is a constant, $T_c$ is a critical temperature, and $T$ is the Universe temperature (see \cite{Gergely:2008jr} for details). Therefore, from a phenomenological point of view, our proposal 
$\hat{\lambda}(z)$ could be a obtained from a generalization of E\"{o}tv\"{o}s law, similar to the generalization of Gauss theorem in n-dimensions.

\textit{Observational constraints.-}
To constrain the $n$ parameter for the $\hat{\lambda}(z)$ ansatz, $\Omega_{m0}$, and $h$,
we perform a Bayesian MCMC analysis using the EMCEE Python module \cite{Foreman:2013}. We choose flat priors on
$n:[0,20]$ and $\Omega_{m0}:[0,1]$, Gaussian prior on $\Omega_{b0}:0.02202\pm0.00046$. After we set burn-in steps to achieve the convergence, a chain of $6000$ MCMC steps with $500$ walkers is performed. We use the following observational data:
\begin{itemize}
\item\textit{H(z) measurements:}
We employ the most recent observational Hubble data measured using cosmic chronometers compiled by Maga\~na et al. 2017 and references therein. This sample contains $31$ points in the redshift range 
$0 \leq z \leq 1.965$. We also consider the local value of the Hubble constant $H_0$ given by A. Riess \cite{Riess:2016jrr}.
This last point is used as a Gaussian prior in the Bayesian analysis. These $H(z)$ measurements could be overestimated up to $25\%$, as claimed by \cite{Wei:2017}. There is a tension, up to more than $3\sigma$, between the local measurements of $H_{0}$ and those obtained from the CMB anisotropies \cite{Bernal:2016}.
\item \textit{Type Ia Supernovae (SN Ia):} we choose the full JLA sample by \cite{Betoule:2014} containing 740 observations in the redshift interval $0.01<z<1.2$. Although several systematics sources have been identified in SN Ia analysis \cite{Conley:2007ng,Conley:2011,Scolnic:2013efb,Scolnic:2013xra,Mosher:2014}, these have been already considered in the covariance matrix provided by \cite{Betoule:2014}. 
\item \textit{Baryon acoustic oscillations:}
Referencing \cite{Garcia-Aspeitia:2016kak}, we use the following BAO measurements: $d_{z}\equiv r_s(z_d)/D_V(z)=0.336\pm0.015$ at redshift $z=0.106$ \cite[6dFGS,][]{Beutler2011:6dF},
$d_{z}=(0.0870\pm0.0042, 0.0672\pm0.0031,0.0593\pm0.0020)$ at $z=(0.44,0.6,0.73)$ \cite[WiggleZ,][]{Kazin:2014,Gong:2015},
$d_{z}=0.2239\pm0.0084$ at $z=0.15$ \cite[SDSS-DR7,][]{Ross:2014},
$d_{z}=(0.1181\pm0.0022,0.0726\pm0.0007)$ at $z=(0.32,0.57)$ \cite[BOSS-DR11,][]{Anderson:2014}
and $D_{H}/r_{d}=9.07\pm0.31$ at $z=2.33$ \cite{Bautista:2017}.

\item \textit{CMB distance posteriors from Planck 2015 measurements:}
We use the acoustic scale, $l_{A}=301.787\pm0.089$, the shift parameter, $R=1.7492\pm0.0049$, 
and the decoupling redshift, $z_{*}=1089.99\pm0.29$ obtained for a flat $w$-cold dark matter model \cite{Neveu:2016,Planck:2015XIV}.
Although this method could lead to biased constraints when used in modified gravity models (see discussion in \cite{PhysRevD.95.023524}), 
we choose these data as a first approach. 
\end{itemize}

\begin{figure}
{\includegraphics[width=0.4\textwidth]{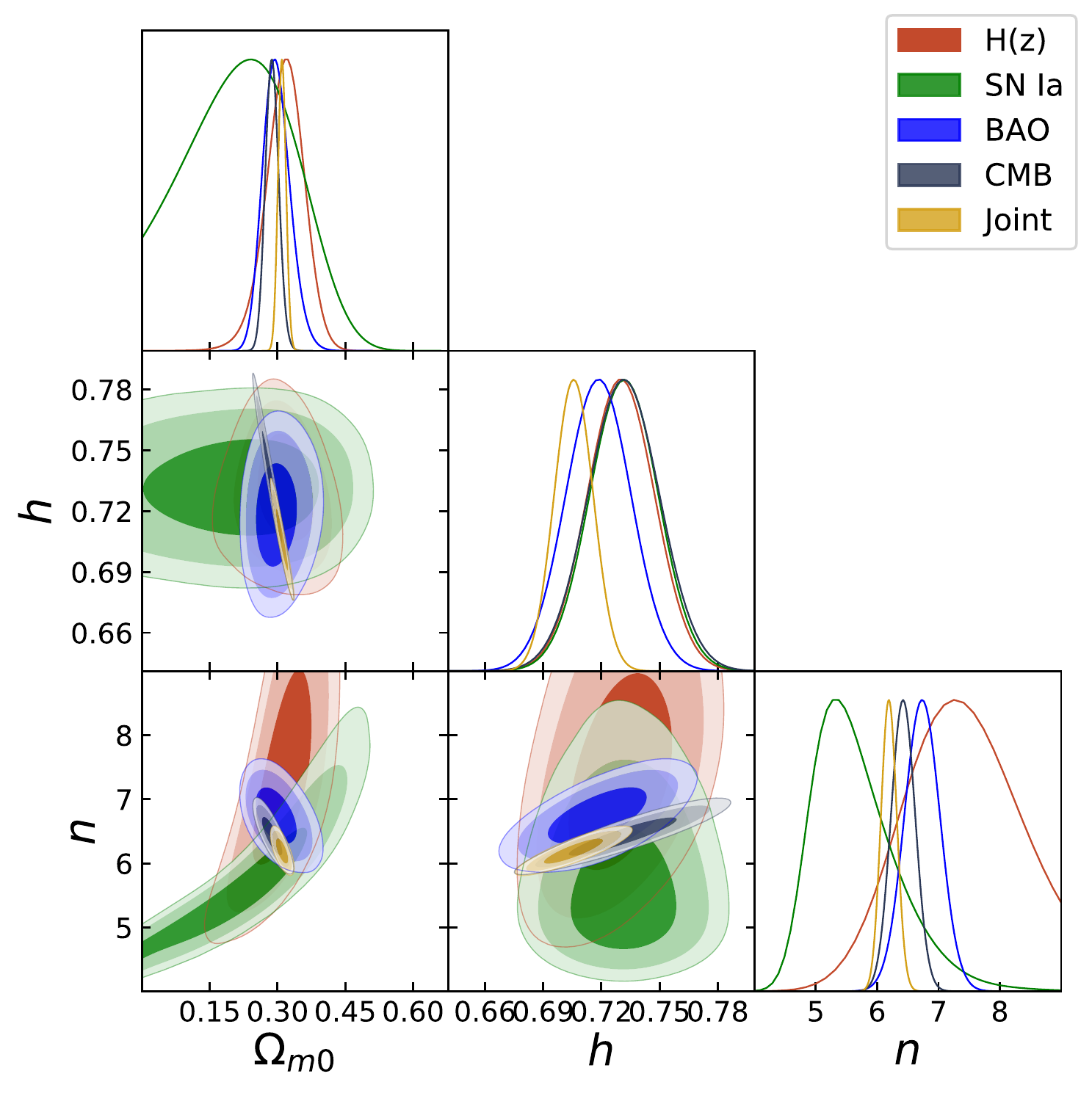}} 
\caption{1D marginalized posterior distributions and the 2D $68\%$, $95\%$, $99.7\%$ of CL for the $\Omega_{m0}$, $h$, and $n$ parameters of the brane model,
assuming a Gaussian prior on $h$ and $\Omega_{b0}$.  }
\label{fig:allcontours}
\end{figure}

Table \ref{tab:par} gives the chi-square and the mean values for the free parameters using each data set. We obtain consistent $\Omega_{m0}$ mean values, within the $1\sigma$ of confidence level (CL), for the different data which are also in agreement with those estimated for the standard scenario. The goodness-of-fit test for the joint analysis indicates that our scenario fits the data with a $95\%$ of reliability. We obtain consistent values for the exponent $n$ within the range $\sim[5.5-7.5]$ at $1\sigma$ CL. Figure \ref{fig:allcontours} presents the $1D$ and $2D$
at $68\%$, $95\%$, $99.7\%$ of CL for the brane parameters. Notice that there is a strong correlation between $n$ and $\Omega_{m0}$ (${\rm corr}(n,\Omega_{m0})=0.912$), i.e. fluctuations on $\Omega_{m0}$ within $1\sigma$ could yield $n$ values larger or smaller than 6. 

\begin{figure}
\includegraphics[width=5cm,scale=0.5]{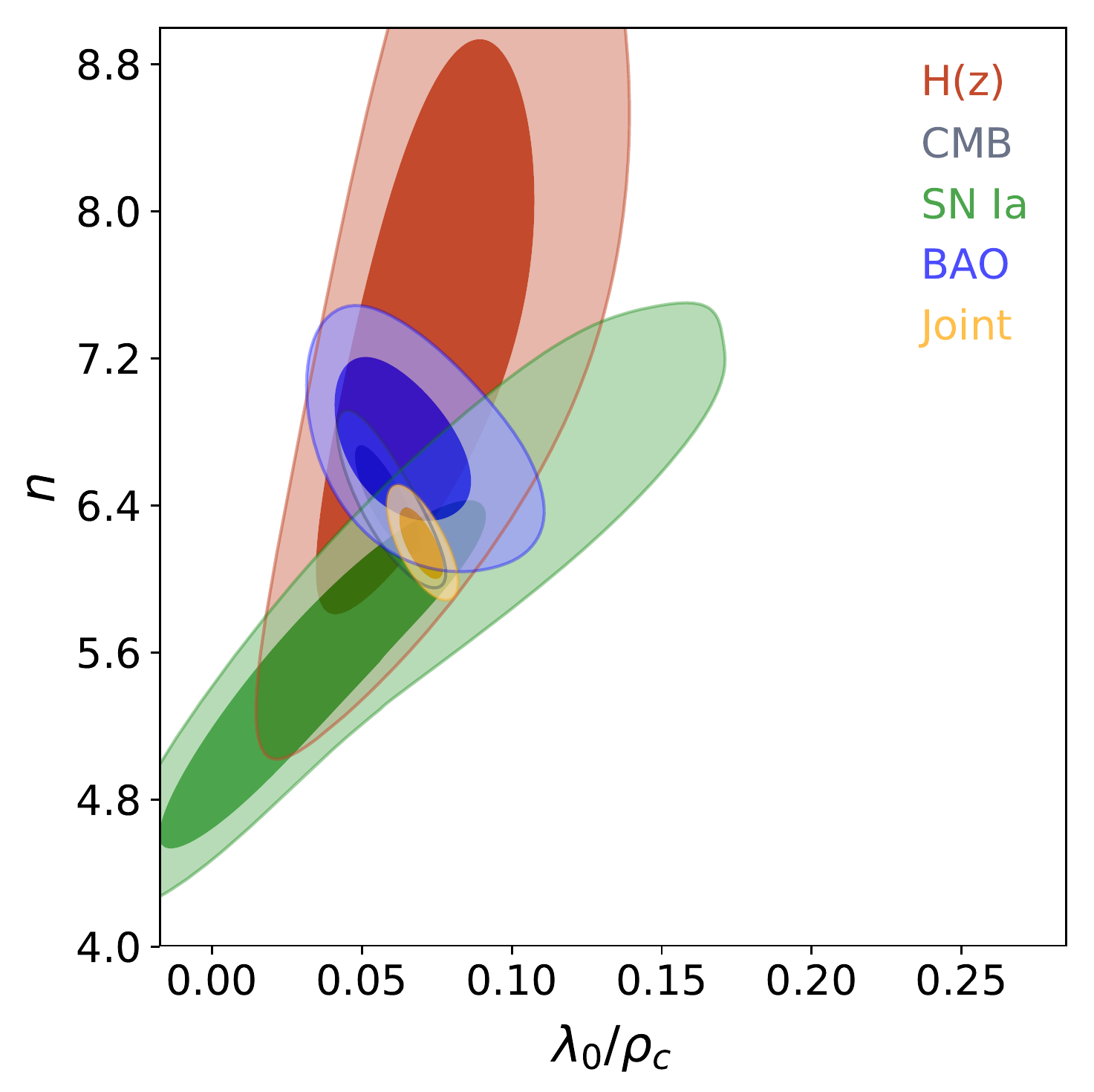} 
\caption{Confidence contours of the $n$-$\lambda_{0}/ \rho_{c}$ 
parameters within the $1\sigma$ and $3\sigma$ of CL for each cosmological data and where $\rho_c=8.070\times10^{-11}h^2$eV$^{4}$. 
\label{fig:lambdacontours}}
\end{figure}

Figure \ref{fig:lambdacontours}
shows the $68\%$ and $99.7\%$ $\lambda_{0}/\rho_{c}$-$n$ confidence contours. Notice that
they overlap at $\lambda_{0}/\rho_{c}\sim0.06$, confirming that the $\lambda_{0}$ constrains are consistent among them, solving the tension between the observables found in \cite{Garcia-Aspeitia:2016kak}, where $\lambda$ was considered constant. Furthermore, the joint analysis value corresponds to a brane tension of the order $\sim10^{40}$eV$^4$ in the Nucleosynthesis epoch ($z_{nuc}\sim 3\times 10^{8}$, $T_{nuc}\sim0.1$MeV), in concordance with Astrophysical observations \cite{Germani:2001du} and previous Nucleosynthesis bounds \cite{Maartens:2003tw}. In addition, it is possible to infer from Eq. \eqref{eq:Momegas} a strong positive correlation of $\lambda_{0}-\Omega_{m}$.

\begin{figure}
\includegraphics[width=5.5cm,scale=0.6]{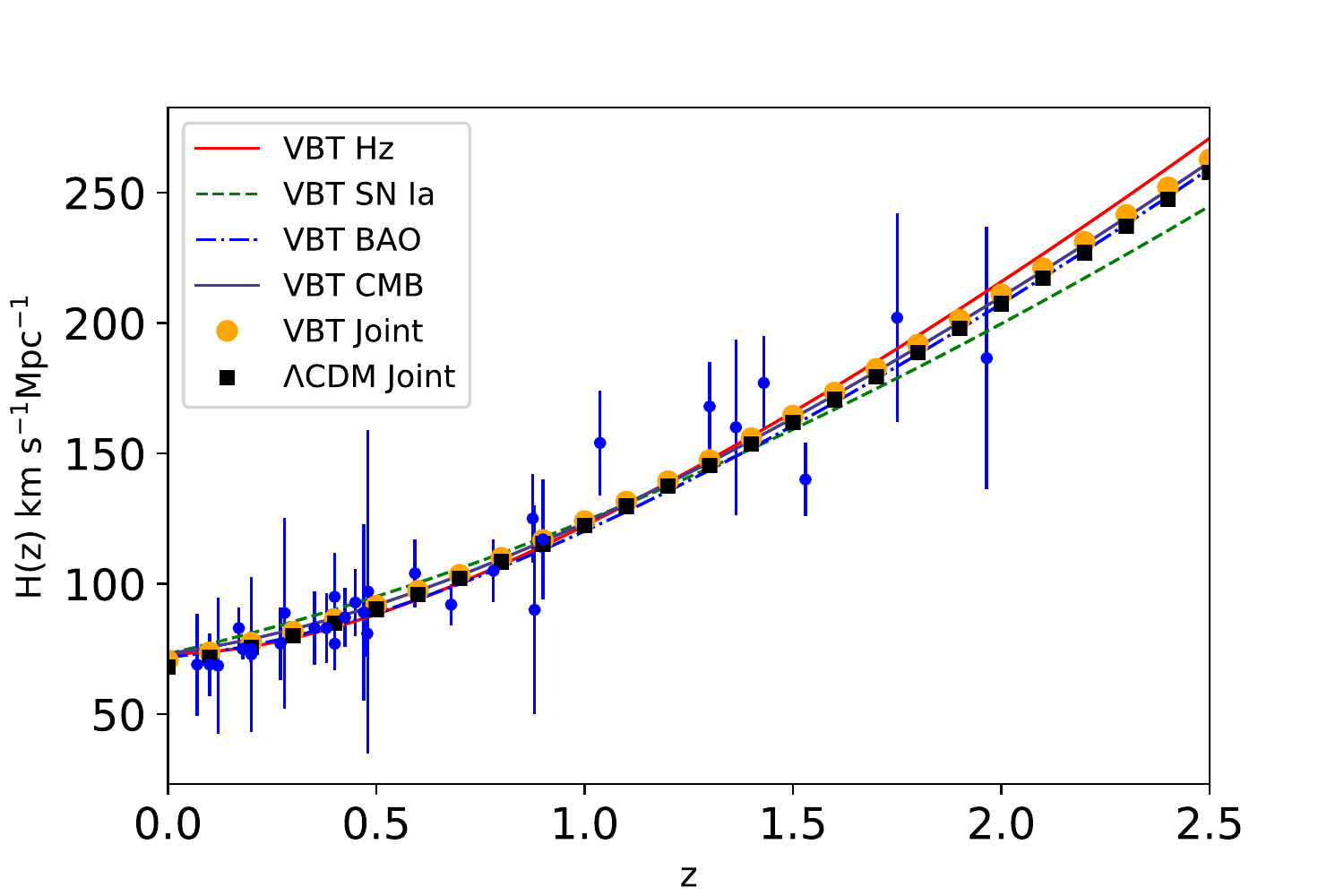}
\includegraphics[width=5.5cm,scale=0.6]{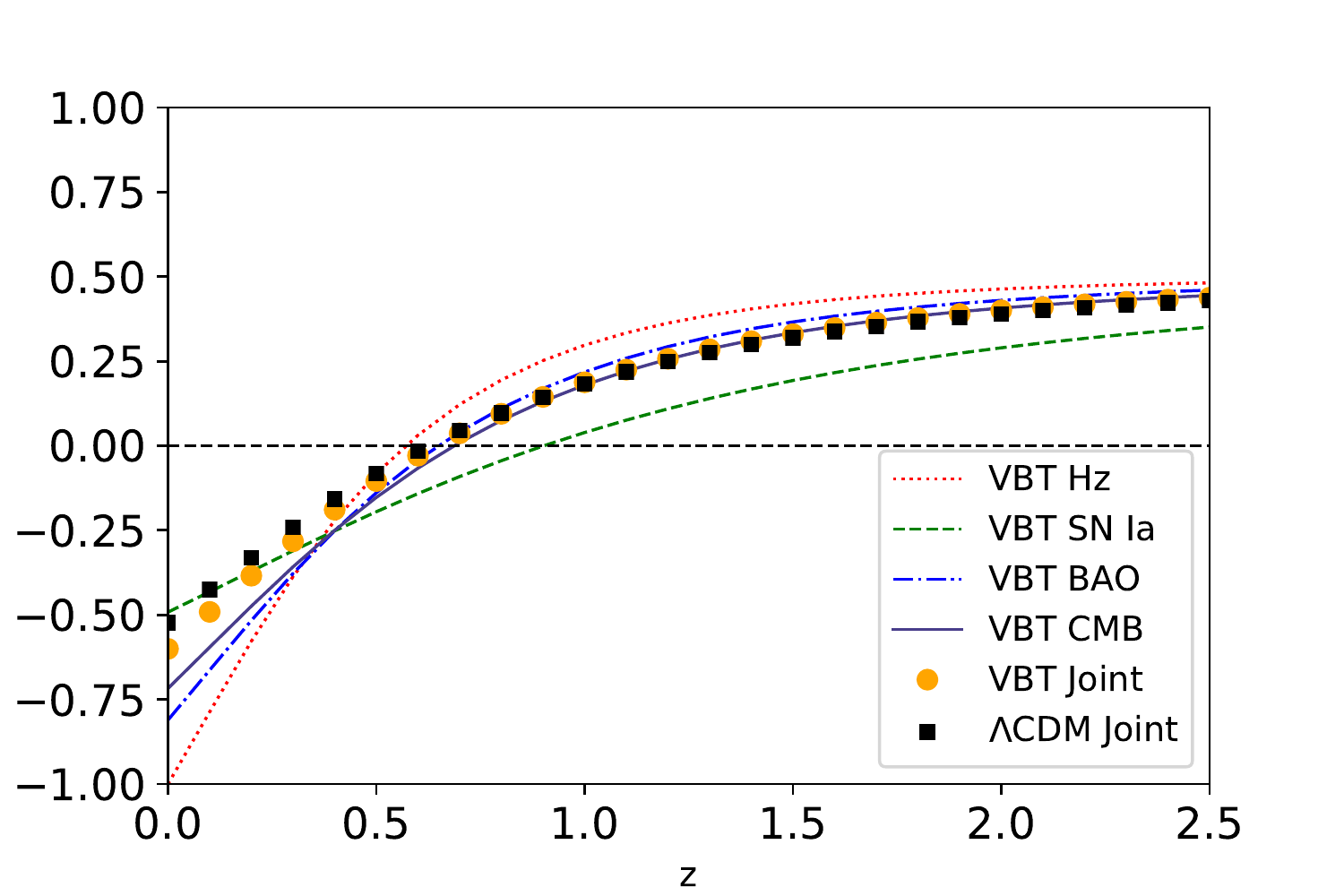} 
\caption{The chrono-brane fitting to H(z) and reconstruction of the deceleration parameter $q(z)$ (top and bottom panels respectively) using the H(z), SN Ia, BAO, CMB, and joint constraints.
The $\Lambda$CDM dynamics also has been plotted for comparison.}
\label{fig:qz}
\end{figure}

Figure \ref{fig:qz} illustrates a good fit to H(z) (top panel) and the reconstructed $q(z)$ (bottom panel) for the chrono-brane model using the mean values for each data set and the joint analysis.
From the latter, we obtain that the Universe starts an accelerated stage at redshift $0.641 \pm 0.018$.
In addition, the $q(z)$ behavior for the chrono-brane model is consistent with the CC dynamics within $1\sigma$ of CL, where for $z=0$ we have $q(0)\simeq-0.60$.

We compare the Akaike information criterion (AIC) \cite{Akaike:1974,Shi:2012ma} and the Bayesian information criterion (BIC) \cite{Schwarz:1978,Shi:2012ma}  between chrono-brane and $\Lambda$CDM models using each dataset.
When the joint constraints are considered, we obtain
$\Delta$AIC$\sim5.6$ and $\Delta$BIC $\sim 0.95$, i.e, weak evidence in favor and not enough evidence against of chrono-brane model. Also, we obtain a Bayes factor \cite{Jarosz:2014} of $1.6$ that gives a weak support of chrono-brane model over standard model as well.

Notice that a value $n=6$ solves the late cosmic acceleration problem without a DE entity. When $\hat{\lambda}(z)=(z+1)^{6}$, the second term in the right-hand-side of Eq. \eqref{EI}
results in $(1-\Omega_{m0}-\Omega_{r0})[\Omega_{m0}^{2}+\Omega_{r0}^{2}(z+1)^{2}]/(\Omega_{m0}^2+\Omega_{r0}^2)$. Since $\Omega_{r0}\sim10^{-5}$, it can be approximated as $\sim(1-\Omega_{m0}-\Omega_{r0})$, same as
for the cosmological constant. 
As a main conclusion, taking into account that the data prefer constraints on $n$ consistent with $n=6$, we suggest that \textit{a brane with variable tension $\lambda(z)=\lambda_{0}(1+z)^{6.19 \pm 0.12}$ can mimic the DE dynamics}. 
Although at first glance this result seems trivial, the origin of the acceleration is different to the one in the standard scenario. In extra dimensional models the topology could influence the acceleration, obtaining in some cases phantom-like dark energy. Our results are also consistent with those explored in \cite{Zhao:2017cud}, where a time-evolution of the dark energy fluid is found.

To take into account the effect of systematics on our constraints, we obtain that the $\lambda_0$-$n$ contours from $H(z)$ data shifts towards smaller values of $n$ (with a best-fit of $6.17^{+ 1.00}_{-0.83}$) and larger values of $\lambda_{0}$ but they are still consistent at $3\sigma$. In the SNIa analysis, the difference in the $\lambda_0$-$n$ confidence contours is negligible. A deviation of $3\%$ on the $n$ ($6.00^{+0.10}_{-0.09}$) value and a shift down of the contours was found in the joint analysis. We also consider smaller errors ($0.75\%$ the original ones) on the $H(z)$ measurements, as suggested by \cite{Wei:2017}, yielding to smaller confidence contours and a deviation of $1\%$ ($7.32^{+0.82}_{- 0.73}$) to our value of $n$. We also use different priors on the SNIa parameters.
The SNIa constraints on $n$ are mainly affected by the $\Omega_{m0}$ estimation. Therefore, although the different systematics in the data introduce different bias in the estimated constraints, the final results are all consistent within the $3\sigma$ CL.
\begin{table}
\caption{Mean values for the brane model parameters ($\Omega_{m0}$, $h$, $n$) derived from H(z), SN Ia, BAO, CMB measurements and a joint analysis.}
\resizebox{0.5\textwidth}{!}{
\begin{tabular}{|cccccc|}
\multicolumn{6}{c}{}\\
\hline
Data set & $\chi^{2}_{min}$& $\Omega_{m0}$ & $h$ &$n$& $\lambda_{0}$($10^{-12}\mathrm{eV}^{4}$)\\
\hline
\multicolumn{6}{|c|}{}\\
H(z) &  14.46  & $0.318^{+0.039}_{-0.042}$ & $0.730^{+0.017}_{-0.017}$ & $7.400^{+1.100}_{-0.926}$&$3.20^{+1.05}_{-0.95}$\\
BAO &  9.49  & $0.297^{+0.031}_{-0.028}$ & $0.718^{+0.016}_{-0.016}$ & $6.730^{+0.287}_{-0.289}$&$2.62^{+0.77}_{-0.57}$\\
CMB & 3.64&  $0.288^{+0.014}_{-0.013}$ & $0.732^{+0.017}_{-0.017}$ & $6.420^{+0.185}_{-0.185}$&$2.52^{+0.19}_{-0.17}$\\
SN Ia & 691.10 & $0.231^{+0.114}_{-0.120}$ & $0.731^{+0.017}_{-0.017}$ & $5.580^{+0.815}_{-0.568}$&$1.48^{+2.40}_{-1.16}$\\
Joint &  716.43 & $0.31^{+0.008}_{-0.008}$ & $0.706^{+0.009}_{-0.009}$ & $6.190^{+0.121}_{-0.120}$&$2.81^{+0.12}_{-0.11}$ \\
\hline
\end{tabular}}
\label{tab:par}
\end{table}

\begin{figure}
\includegraphics[width=5.5cm,scale=0.5]{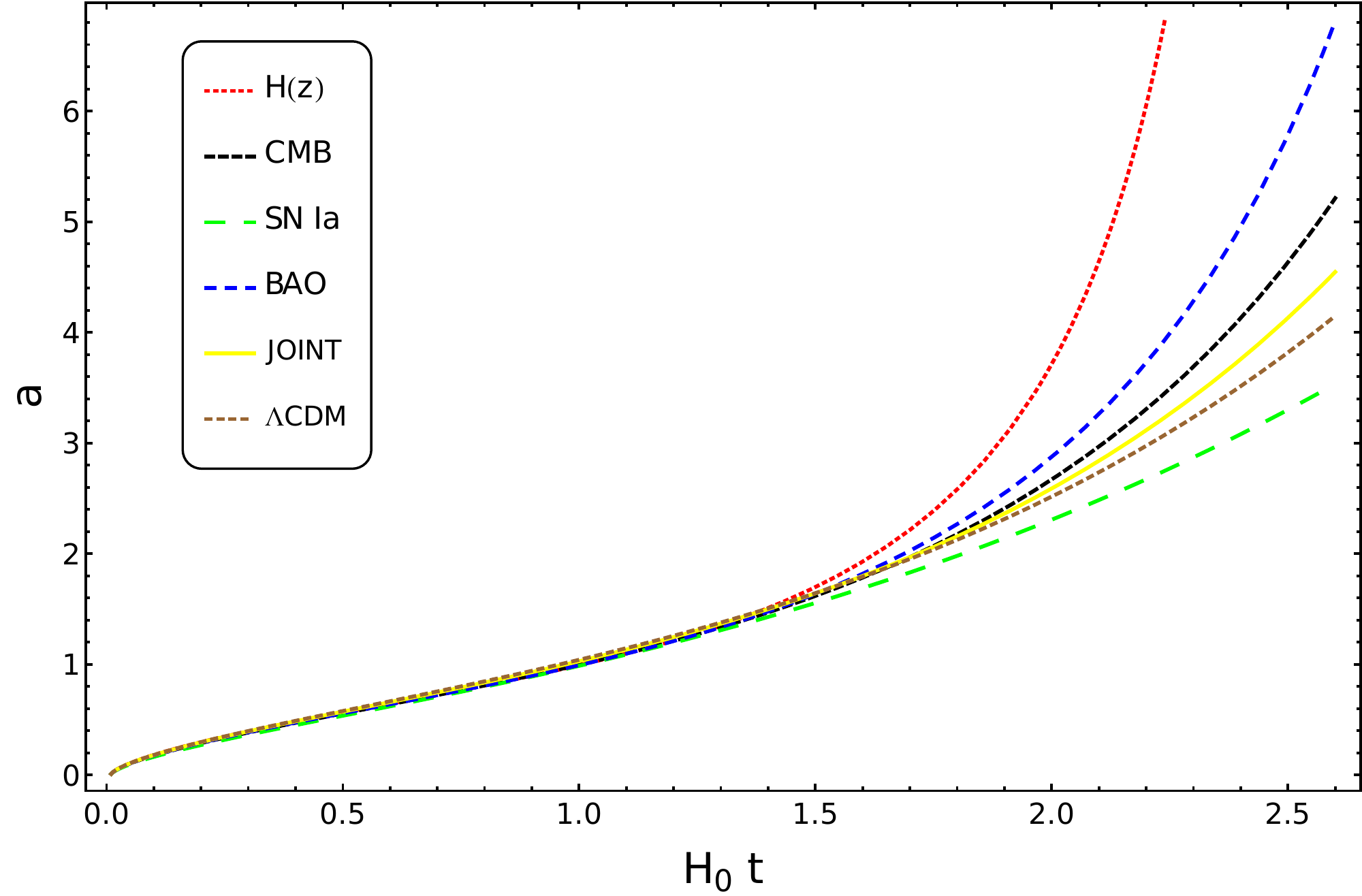}
\caption{Evolution of the scale factor (Eq. \eqref{A0I}), assuming a non singular initial condition as $a(10^{-2})=10^{-2}$, using H(z), BAO, CMB, SNIa and joint constraints. A comparison with $\Lambda$CDM it is also shown.}
\label{sf}
\end{figure}

\textit{Scale factor dynamics.-} 
In order to study the scale factor dynamics, the Friedmann equation can be written in terms of quadratures as:
\begin{equation}
H_0(t-t_0) = \int_{a_0}^a\left\lbrace\frac{\Omega _{m0}}{a}+ \frac{\Omega _{r0}}{a^2} 
+ \mathcal{M}\left[\frac{\Omega _{m0}^2}{a^{4-n}} + \frac{\Omega _{r0}^2}{a^{6-n}} \right] \right\rbrace^{-1/2}da.\label{A0I}
\end{equation}
The numerical solution of the Eq. \eqref{A0I} using the mean value constraints is shown in Fig. \ref{sf}. 
In this case, we have assumed a non singular initial condition $a(10^{-2})=10^{-2}$, i.e., 
this model presents a Big Bang singularity, in concordance with the traditional models. 
In contrast, it is possible to observe late times singularities for the values of $n$, $\Omega_{m0}$, and $h$ shown in Table \ref{tab:par}.
Future singularities at $t_{sing}$ times, can be computed by $t_{sing}-t_{today}\simeq2H_0^{-1}(n-6)^{-1}\mathcal{M}^{-1/2}\Omega_{0m}^{-1}$. 
Particularly, we observe singularities for the constraints obtained from H(z) at $t_{sing}=2.69H_0^{-1}$, BAO at $t_{sing}=4.22H_0^{-1}$, CMB at $t_{sing}=6.58H_0^{-1}$ and from the joint  analysis we have $t_{sing}=13.57H_0^{-1}$; implying that the fate of the Universe is a Big Rip, i.e. behaves as phantom-like in traditional general relativity.
Notice that constraints relying only in SN Ia analysis do not predict future singularities. 

The approximate analytical solution of the scale factor as a function of time reads as
\begin{equation}
     \label{SFL}
     a(t) \simeq \left\{
	       \begin{array}{ll}
		 \left[(3-\frac{n}{2})\alpha\Delta t+a_0^{(6-n)/2}\right]^{2/(6-n)},     & \mathrm{for\ } n\neq6, \\
  a_0\exp(\alpha\Delta t),   & \mathrm{for\ } n=6,
	       \end{array}
	     \right.
   \end{equation}
where $\alpha\equiv\Omega_{m0}\mathcal{M}^{1/2}H_0$ and $\Delta t\equiv t-t_0$. Notice that the d'Sitter expansion in the second case, which it is straightforward from Eq. \eqref{EI}, behaves like a cosmological constant. 

An effective equation of state (EoS) is calculated from the Friedmann and Raychaudhuri equations as
\begin{eqnarray}
\omega_{eff}(z)&=&\frac{2q(z)-1}{3[1+2\mathcal{M}E(z)^2_{nb}(z+1)^{-n}]}+\nonumber\\&&\frac{\mathcal{M} E(z)^2_{nb}(z+1)^{-n} [2q(z)-(4-n)]}{3[1+2\mathcal{M}E(z)^2_{nb}(z+1)^{-n}]}. \label{eos}
\end{eqnarray}
The Universe accelerates when $\omega_{eff}$ satisfies
\begin{equation}
\omega_{eff}(z)<-\frac{1+\mathcal{M} E(z)^2_{nb}(z+1)^{-n} (4-n)}{3[1+2\mathcal{M}E(z)^2_{nb}(z+1)^{-n}]}. \label{wmen}
\end{equation}
The reader must be notice that the $\omega_{eff}(z)$ from GR is not valid anymore in this particular case.

\begin{figure}
\includegraphics[width=6.4cm,height=4.5cm]{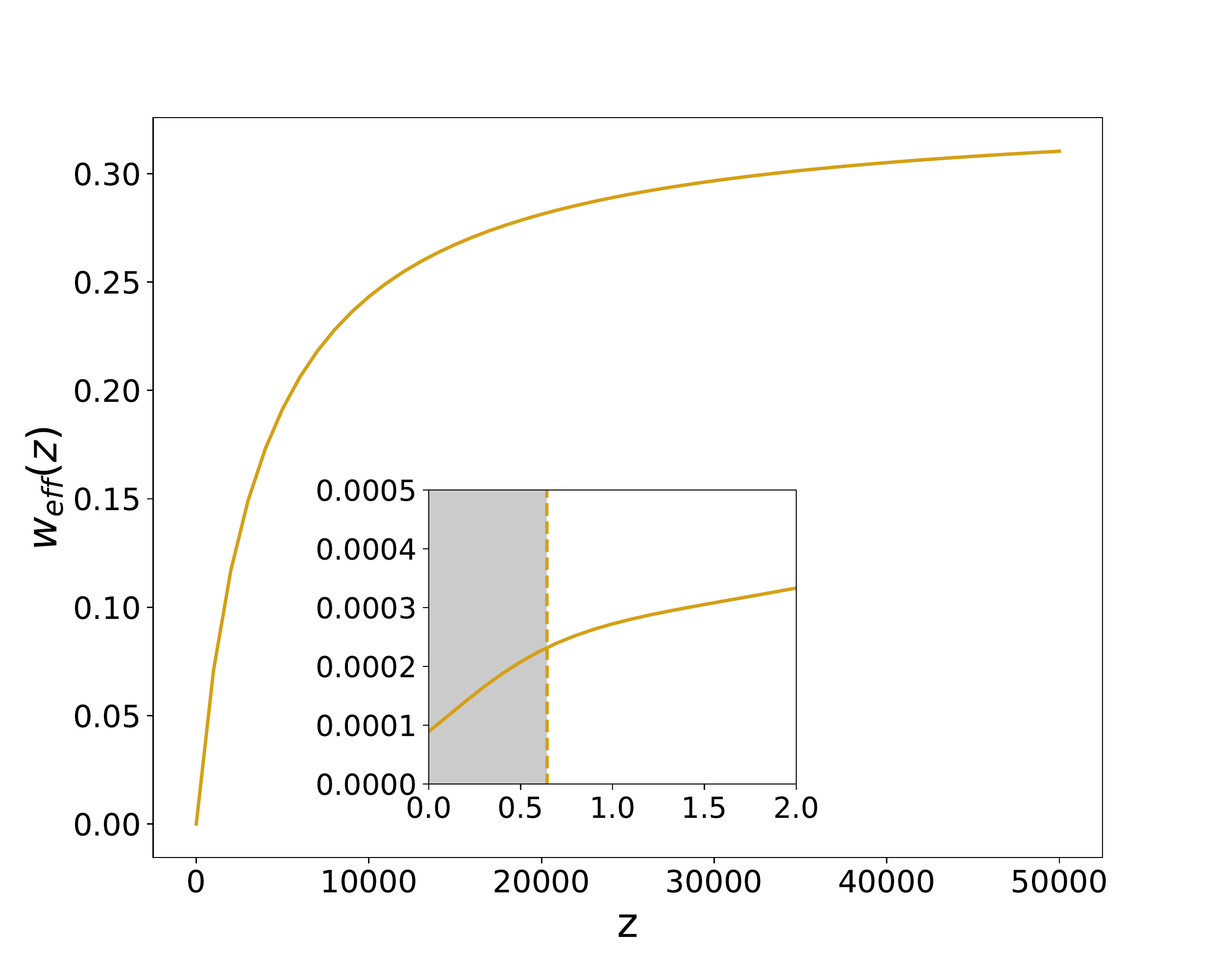}
\caption{Reconstruction of the effective EoS using the joint constraints. The inset shows the current $\omega_{eff}$ behavior. The vertical dashed line marks the redshift where the condition of Eq. \eqref{wmen} is satisfied, i.e. the Universe accelerates for lower redshifts than $0.65$.}
\label{fig:eos}
\end{figure}
Figure \ref{fig:eos} shows the effective Eos evolution using the joint constraints. Notice that 
$\omega_{eff}\rightarrow1/3$ at high redshifts and $\omega_{eff}\rightarrow0$ at $z=0$. The inset shows the region where the condition
(\ref{wmen}) is satisfied (at $z\lesssim0.65$ when $w_{eff}<0.00025$), i.e., when the Universe accelerates. Notice that the transition redshift in the $w_{eff}$ is consistent
with the one obtained in the $q(z)$ reconstruction.

\begin{figure}
\includegraphics[width=5.6cm,scale=0.6]{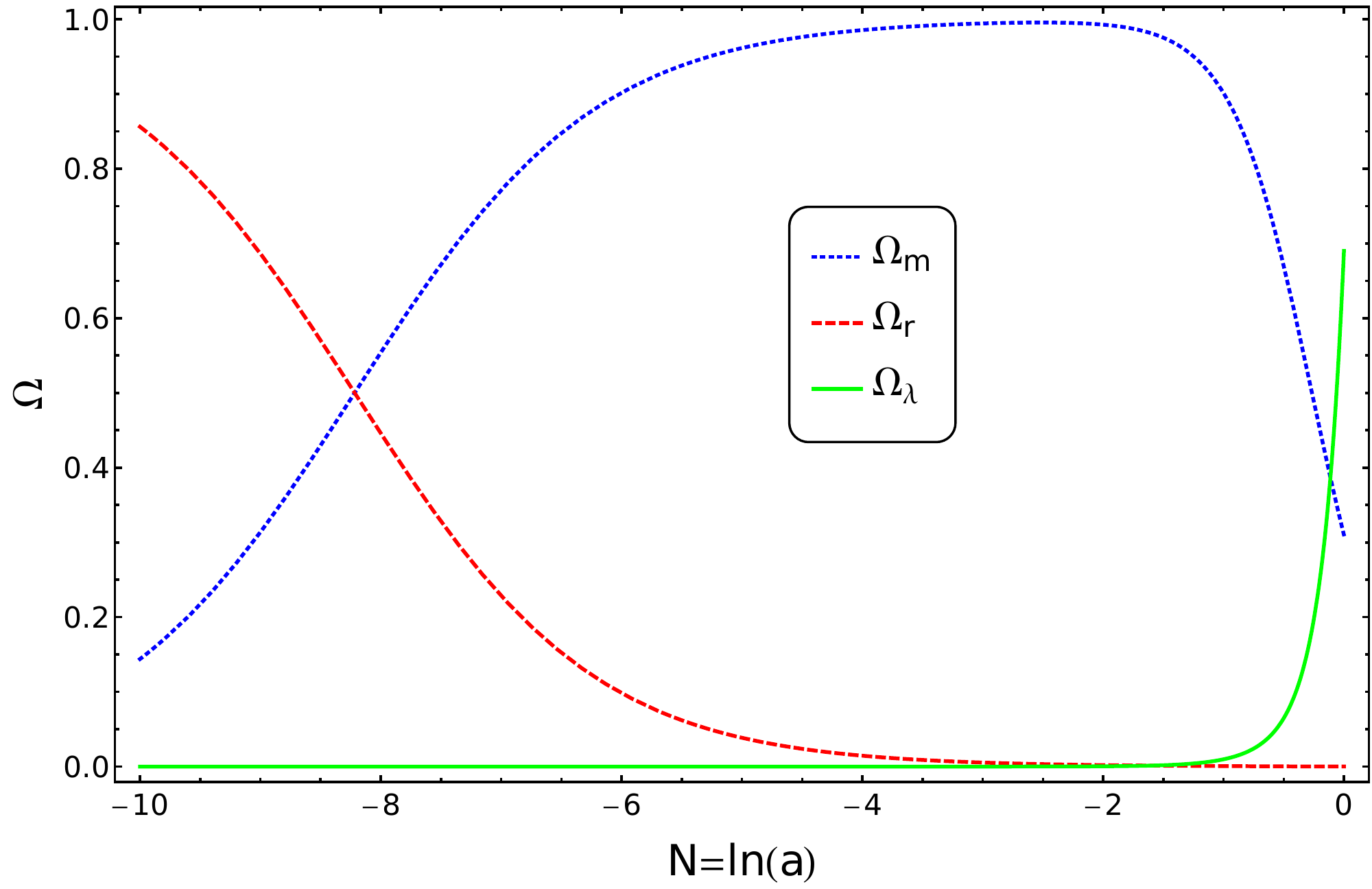}
\caption{Evolution of the density parameters under the chrono-brane scenario. The initial conditions are chosen from the mean values of joint constraints shown in Table \ref{tab:par}.}
\label{ds}
\end{figure}

\textit{Cosmological evolution.- }
Proposing the following dimensionless variables:
\begin{eqnarray}
&&x^2\equiv\Omega_m=\left(\frac{\kappa^2_{(4)}\rho_m}{3H^2}\right), \qquad y^2\equiv\Omega_r=\left(\frac{\kappa^2_{(4)}\rho_r}{3H^2}\right), \qquad \nonumber\\
&&z^2\equiv\frac{3H^2}{2\kappa^2_{(4)}\lambda_0\bar{\lambda}(a)}, \label{dyneq0}
\end{eqnarray}
where $\Omega_{\lambda}=z^2(x^4+y^4)=1-\Omega_m-\Omega_r$, with the Friedmann constraint
$1=x^2+y^2+z^2(x^4+y^4)$, allows us to construct the following dynamical system

\begin{subequations}
\begin{eqnarray}
&&\frac{x^{\prime}}{x}=-\frac{3}{2}+\frac{3}{2}x^2+2y^2-\frac{1}{2}\Pi, \\
&&\frac{y^{\prime}}{y}=-2+\frac{3}{2}x^2+2y^2-\frac{1}{2}\Pi, \\
&&\frac{z^{\prime}}{z}=\frac{n}{2}-\frac{3}{2}x^2-2y^2+\frac{1}{2}\Pi,
\end{eqnarray}
\end{subequations}
where $\Pi\equiv\left[(n-6)x^4+(n-8)y^4\right]z^2$. Choosing as initial condition the Joint constraints and numerically solving this dynamical system, we obtain the evolution of the density parameters shown in Fig. \ref{ds}. At early times, the Universe is dominated by the radiation component, after, the matter becomes the dominant component. At late times, the Universe is dominated by the chrono-brane dynamics. Therefore, this scenario predicts the same cosmological evolution as the standard one.

\textit{Conclusions and Discussion.-}
In this letter we constructed a brane world model which produce an accelerated Universe without a dark energy component. 
We present a new way of building RS models using a variable brane tension $\lambda(z)$, called chrono-brane. 
We introduce the ansatz $\lambda(z)=(z+1)^{n}$, inspired by tracker-scalar-field potentials, arising from the space-time structure. To constrain the $n$ exponent, the matter content, and 
the dimensionless Hubble parameters we used H(z), SNIa, BAO and CMB cosmological observations. We found consistent mean values
for the different parameters using each set of observational data. From the joint analysis we estimated \textbf{$n\sim6.19 \pm 0.12$}, i.e. the data prefer a $n$ value providing a term in the Friedmann equation which mimics the DE dynamics very close to a CC at late times. In addition, $\Omega_{m0}$ and $h$ are in excellent agreement with the standard values. Our model also alleviates the tension among the $\lambda_{0}$ constraints obtained from the cosmological data and those from high-energy-regime. For example, we obtain from the joint analysis $\lambda= 8.35\times10^{40}\mathrm{eV}^4$ at $z\sim 3\times10^{8}$ for Nucleosynthesis epoch, that would not affect well-established primordial processes. For the current epoch $z=0$, we have $\lambda=2.81\times 10^{-12}\mathrm{eV}^{4}$.

All of our cosmological constraints give a good fit to H(z) and predict a phase of accelerated expansion at $z\sim 0.6$. Our results on the scale factor evolution exhibits a future singularity, i.e. the fate of the Universe is a Big Rip, as it also happens with phantom DE. We reconstructed the cosmological behavior of an effective EoS and found that the Universe accelerates when $\omega_{eff} <0.00025$ at $z<0.65$, obtaining $q(0)\simeq-0.60$. 
We studied the density parameter evolution for each component and recovered a value that is the same as the standard one. This is a key result because a chrono-brane successfully reproduces the concordance model and provides clues to the DE nature and the late cosmic acceleration. 

Further analysis of the brane perturbations would give information about the viability of chrono-branes. In this vein, Ref. \cite{Koyama:2003be} explore the consequences of a simple brane model with constant brane tension on the CMB spectrum. The authors show that at large scales 
the temperature anisotropy caused by Sachs-Wolfe effect is the same as the canonical one. They also claim that at very small scales the effects of branes are negligible. Nevertheless, on scales up to the first CMB acoustic peak,  the brane terms considerably modify the peak amplitude and position. This implies a change in the CMB distance posteriors and, thus, in the brane constraints that we have obtained. It is important to notice that these results are also applicable for the case of constant brane tension. However, to asses the impact of the perturbation on the brane constraints, a full CMB analysis should be carried out, which is beyond of the scope of this article.

\begin{acknowledgments}
\textit{Acknowledgments.-} We thank the anonymous referee for thoughtful remarks and suggestions. C.M.-R. and M.H.A. acknowledge the support provided by CONACyT fellowship; 
M.H.A. also acknowledges Centro de Astrof\'{\i}sica de Valpara\'{\i}so (CAV) and 
thanks the staff of the Instituto de F\'{\i}sica y Astronom\'{\i}a of the Universidad de Valpara\'{\i}so where part of this work was done.
M.A.G.-A. acknowledges support from SNI-M\'exico and CONACyT research fellow. J.M. acknowledges support from FONDECYT 3160674. A.H.A. acknowledges support from SNI-M\'exico; 
Instituto Avanzado de Cosmolog\'ia (IAC) collaborations. The authors thankfully acknowledge computer resources, technical advise and support provided by Laboratorio de Matem\'atica Aplicada y Computo de Alto Rendimiento del CINVESTAV-IPN (ABACUS), Proyecto CONACYT-EDOMEX-2011-C01-165873.\\
\end{acknowledgments}

\bibliography{librero1}

\begin{thebibliography}{53}%
\makeatletter
\providecommand \@ifxundefined [1]{%
 \@ifx{#1\undefined}
}%
\providecommand \@ifnum [1]{%
 \ifnum #1\expandafter \@firstoftwo
 \else \expandafter \@secondoftwo
 \fi
}%
\providecommand \@ifx [1]{%
 \ifx #1\expandafter \@firstoftwo
 \else \expandafter \@secondoftwo
 \fi
}%
\providecommand \natexlab [1]{#1}%
\providecommand \enquote  [1]{``#1''}%
\providecommand \bibnamefont  [1]{#1}%
\providecommand \bibfnamefont [1]{#1}%
\providecommand \citenamefont [1]{#1}%
\providecommand \href@noop [0]{\@secondoftwo}%
\providecommand \href [0]{\begingroup \@sanitize@url \@href}%
\providecommand \@href[1]{\@@startlink{#1}\@@href}%
\providecommand \@@href[1]{\endgroup#1\@@endlink}%
\providecommand \@sanitize@url [0]{\catcode `\\12\catcode `\$12\catcode
  `\&12\catcode `\#12\catcode `\^12\catcode `\_12\catcode `\%12\relax}%
\providecommand \@@startlink[1]{}%
\providecommand \@@endlink[0]{}%
\providecommand \url  [0]{\begingroup\@sanitize@url \@url }%
\providecommand \@url [1]{\endgroup\@href {#1}{\urlprefix }}%
\providecommand \urlprefix  [0]{URL }%
\providecommand \Eprint [0]{\href }%
\providecommand \doibase [0]{http://dx.doi.org/}%
\providecommand \selectlanguage [0]{\@gobble}%
\providecommand \bibinfo  [0]{\@secondoftwo}%
\providecommand \bibfield  [0]{\@secondoftwo}%
\providecommand \translation [1]{[#1]}%
\providecommand \BibitemOpen [0]{}%
\providecommand \bibitemStop [0]{}%
\providecommand \bibitemNoStop [0]{.\EOS\space}%
\providecommand \EOS [0]{\spacefactor3000\relax}%
\providecommand \BibitemShut  [1]{\csname bibitem#1\endcsname}%
\let\auto@bib@innerbib\@empty
\bibitem [{\citenamefont {Schmidt}\ \emph {et~al.}(1998)\citenamefont
  {Schmidt}, \citenamefont {Suntzeff}, \citenamefont {Phillips}, \citenamefont
  {Schommer}, \citenamefont {Clocchiatti} \emph {et~al.}}]{Schmidt}%
  \BibitemOpen
  \bibfield  {author} {\bibinfo {author} {\bibfnamefont {B.~P.}\ \bibnamefont
  {Schmidt}}, \bibinfo {author} {\bibfnamefont {N.~B.}\ \bibnamefont
  {Suntzeff}}, \bibinfo {author} {\bibfnamefont {M.~M.}\ \bibnamefont
  {Phillips}}, \bibinfo {author} {\bibfnamefont {R.~A.}\ \bibnamefont
  {Schommer}}, \bibinfo {author} {\bibfnamefont {A.}~\bibnamefont
  {Clocchiatti}},  \emph {et~al.},\ }\href
  {http://stacks.iop.org/0004-637X/507/i=1/a=46} {\bibfield  {journal}
  {\bibinfo  {journal} {The Astrophysical Journal}\ }\textbf {\bibinfo {volume}
  {507}},\ \bibinfo {pages} {46} (\bibinfo {year} {1998})}\BibitemShut
  {NoStop}%
\bibitem [{\citenamefont {Perlmutter}\ \emph {et~al.}(1999)\citenamefont
  {Perlmutter}, \citenamefont {Aldering}, \citenamefont {Goldhaber},
  \citenamefont {Knop}, \citenamefont {Nugent}, \citenamefont {others},\ and\
  \citenamefont {Project}}]{Perlmutter}%
  \BibitemOpen
  \bibfield  {author} {\bibinfo {author} {\bibfnamefont {S.}~\bibnamefont
  {Perlmutter}}, \bibinfo {author} {\bibfnamefont {G.}~\bibnamefont
  {Aldering}}, \bibinfo {author} {\bibfnamefont {G.}~\bibnamefont {Goldhaber}},
  \bibinfo {author} {\bibfnamefont {R.~A.}\ \bibnamefont {Knop}}, \bibinfo
  {author} {\bibfnamefont {P.}~\bibnamefont {Nugent}}, \bibinfo {author}
  {\bibnamefont {others}}, \ and\ \bibinfo {author} {\bibfnamefont {T.~S.~C.}\
  \bibnamefont {Project}},\ }\href
  {http://stacks.iop.org/0004-637X/517/i=2/a=565} {\bibfield  {journal}
  {\bibinfo  {journal} {The Astrophysical Journal}\ }\textbf {\bibinfo {volume}
  {517}},\ \bibinfo {pages} {565} (\bibinfo {year} {1999})}\BibitemShut
  {NoStop}%
\bibitem [{\citenamefont {Riess}\ \emph {et~al.}(1998)\citenamefont {Riess},
  \citenamefont {Filippenko}, \citenamefont {Challis}, \citenamefont
  {Clocchiatti}, \citenamefont {Diercks} \emph {et~al.}}]{Riess}%
  \BibitemOpen
  \bibfield  {author} {\bibinfo {author} {\bibfnamefont {A.~G.}\ \bibnamefont
  {Riess}}, \bibinfo {author} {\bibfnamefont {A.~V.}\ \bibnamefont
  {Filippenko}}, \bibinfo {author} {\bibfnamefont {P.}~\bibnamefont {Challis}},
  \bibinfo {author} {\bibfnamefont {A.}~\bibnamefont {Clocchiatti}}, \bibinfo
  {author} {\bibfnamefont {A.}~\bibnamefont {Diercks}},  \emph {et~al.},\
  }\href {http://stacks.iop.org/1538-3881/116/i=3/a=1009} {\bibfield  {journal}
  {\bibinfo  {journal} {The Astronomical Journal}\ }\textbf {\bibinfo {volume}
  {116}},\ \bibinfo {pages} {1009} (\bibinfo {year} {1998})}\BibitemShut
  {NoStop}%
\bibitem [{\citenamefont {Ade}\ \emph {et~al.}(2015{\natexlab{a}})\citenamefont
  {Ade} \emph {et~al.}}]{Planck:2015XIII}%
  \BibitemOpen
  \bibfield  {author} {\bibinfo {author} {\bibfnamefont {P.~A.~R.}\
  \bibnamefont {Ade}} \emph {et~al.} (\bibinfo {collaboration} {Planck}),\
  }\href@noop {} {\  (\bibinfo {year} {2015}{\natexlab{a}})},\ \Eprint
  {http://arxiv.org/abs/1502.01589} {arXiv:1502.01589 [astro-ph.CO]}
  \BibitemShut {NoStop}%
\bibitem [{\citenamefont {Ade}\ \emph {et~al.}(2015{\natexlab{b}})\citenamefont
  {Ade} \emph {et~al.}}]{Planck:2015XIV}%
  \BibitemOpen
  \bibfield  {author} {\bibinfo {author} {\bibfnamefont {P.~A.~R.}\
  \bibnamefont {Ade}} \emph {et~al.} (\bibinfo {collaboration} {Planck}),\
  }\href@noop {} {\  (\bibinfo {year} {2015}{\natexlab{b}})},\ \Eprint
  {http://arxiv.org/abs/1502.01590} {arXiv:1502.01590 [astro-ph.CO]}
  \BibitemShut {NoStop}%
\bibitem [{\citenamefont {Alam}\ \emph
  {et~al.}(2017{\natexlab{a}})\citenamefont {Alam} \emph {et~al.}}]{Alam:2017}%
  \BibitemOpen
  \bibfield  {author} {\bibinfo {author} {\bibfnamefont {S.}~\bibnamefont
  {Alam}} \emph {et~al.} (\bibinfo {collaboration} {BOSS}),\ }\href {\doibase
  10.1093/mnras/stx721} {\bibfield  {journal} {\bibinfo  {journal} {Mon. Not.
  Roy. Astron. Soc.}\ }\textbf {\bibinfo {volume} {470}},\ \bibinfo {pages}
  {2617} (\bibinfo {year} {2017}{\natexlab{a}})},\ \Eprint
  {http://arxiv.org/abs/1607.03155} {arXiv:1607.03155} \BibitemShut {NoStop}%
\bibitem [{\citenamefont {Copeland}\ \emph {et~al.}(2006)\citenamefont
  {Copeland}, \citenamefont {Sami},\ and\ \citenamefont
  {Tsujikawa}}]{Copeland:2006wr}%
  \BibitemOpen
  \bibfield  {author} {\bibinfo {author} {\bibfnamefont {E.~J.}\ \bibnamefont
  {Copeland}}, \bibinfo {author} {\bibfnamefont {M.}~\bibnamefont {Sami}}, \
  and\ \bibinfo {author} {\bibfnamefont {S.}~\bibnamefont {Tsujikawa}},\ }\href
  {\doibase 10.1142/S021827180600942X} {\bibfield  {journal} {\bibinfo
  {journal} {Int. J. Mod. Phys.}\ }\textbf {\bibinfo {volume} {D15}},\ \bibinfo
  {pages} {1753} (\bibinfo {year} {2006})},\ \Eprint
  {http://arxiv.org/abs/hep-th/0603057} {arXiv:hep-th/0603057 [hep-th]}
  \BibitemShut {NoStop}%
\bibitem [{\citenamefont {{Planck Collaboration}}\ \emph
  {et~al.}()\citenamefont {{Planck Collaboration}} \emph
  {et~al.}}]{PlanckCollaboration2013}%
  \BibitemOpen
  \bibfield  {author} {\bibinfo {author} {\bibnamefont {{Planck
  Collaboration}}} \emph {et~al.},\ }\href@noop {} {\bibinfo  {journal}
  {arXiv:1303.5076}\ }\BibitemShut {NoStop}%
\bibitem [{\citenamefont {Weinberg}(1989)}]{Weinberg}%
  \BibitemOpen
\bibfield  {journal} {  }\bibfield  {author} {\bibinfo {author} {\bibfnamefont
  {S.}~\bibnamefont {Weinberg}},\ }\href@noop {} {\bibfield  {journal}
  {\bibinfo  {journal} {Reviews of Modern Physics}\ }\textbf {\bibinfo {volume}
  {61}} (\bibinfo {year} {1989})}\BibitemShut {NoStop}%
\bibitem [{\citenamefont {Zeldovich}(1968)}]{Zeldovich}%
  \BibitemOpen
  \bibfield  {author} {\bibinfo {author} {\bibfnamefont {Y.~B.}\ \bibnamefont
  {Zeldovich}},\ }\href@noop {} {\bibfield  {journal} {\bibinfo  {journal}
  {Soviet Physics Uspekhi}\ }\textbf {\bibinfo {volume} {11}} (\bibinfo {year}
  {1968})}\BibitemShut {NoStop}%
\bibitem [{\citenamefont {Randall}\ and\ \citenamefont
  {Sundrum}(1999{\natexlab{a}})}]{Randall-I}%
  \BibitemOpen
  \bibfield  {author} {\bibinfo {author} {\bibfnamefont {L.}~\bibnamefont
  {Randall}}\ and\ \bibinfo {author} {\bibfnamefont {R.}~\bibnamefont
  {Sundrum}},\ }\href {\doibase 10.1103/PhysRevLett.83.3370} {\bibfield
  {journal} {\bibinfo  {journal} {Phys. Rev. Lett.}\ }\textbf {\bibinfo
  {volume} {83}},\ \bibinfo {pages} {3370} (\bibinfo {year}
  {1999}{\natexlab{a}})},\ \Eprint {http://arxiv.org/abs/hep-ph/9905221}
  {arXiv:hep-ph/9905221} \BibitemShut {NoStop}%
\bibitem [{\citenamefont {Randall}\ and\ \citenamefont
  {Sundrum}(1999{\natexlab{b}})}]{Randall-II}%
  \BibitemOpen
  \bibfield  {author} {\bibinfo {author} {\bibfnamefont {L.}~\bibnamefont
  {Randall}}\ and\ \bibinfo {author} {\bibfnamefont {R.}~\bibnamefont
  {Sundrum}},\ }\href {\doibase 10.1103/PhysRevLett.83.4690} {\bibfield
  {journal} {\bibinfo  {journal} {Phys. Rev. Lett.}\ }\textbf {\bibinfo
  {volume} {83}},\ \bibinfo {pages} {4690} (\bibinfo {year}
  {1999}{\natexlab{b}})},\ \Eprint {http://arxiv.org/abs/hep-th/9906064}
  {arXiv:hep-th/9906064 [hep-th]} \BibitemShut {NoStop}%
\bibitem [{\citenamefont {Dvali}\ \emph {et~al.}(2000)\citenamefont {Dvali},
  \citenamefont {Gabadadze},\ and\ \citenamefont {Porrati}}]{Dvali:2000hr}%
  \BibitemOpen
  \bibfield  {author} {\bibinfo {author} {\bibfnamefont {G.~R.}\ \bibnamefont
  {Dvali}}, \bibinfo {author} {\bibfnamefont {G.}~\bibnamefont {Gabadadze}}, \
  and\ \bibinfo {author} {\bibfnamefont {M.}~\bibnamefont {Porrati}},\ }\href
  {\doibase 10.1016/S0370-2693(00)00669-9} {\bibfield  {journal} {\bibinfo
  {journal} {Phys. Lett.}\ }\textbf {\bibinfo {volume} {B485}},\ \bibinfo
  {pages} {208} (\bibinfo {year} {2000})},\ \Eprint
  {http://arxiv.org/abs/hep-th/0005016} {arXiv:hep-th/0005016 [hep-th]}
  \BibitemShut {NoStop}%
\bibitem [{\citenamefont {Garcia-Aspeitia}\ \emph {et~al.}(2018)\citenamefont
  {Garcia-Aspeitia}, \citenamefont {Maga\~na}, \citenamefont
  {Hernandez-Almada},\ and\ \citenamefont {Motta}}]{Garcia-Aspeitia:2016kak}%
  \BibitemOpen
  \bibfield  {author} {\bibinfo {author} {\bibfnamefont {M.~A.}\ \bibnamefont
  {Garcia-Aspeitia}}, \bibinfo {author} {\bibfnamefont {J.}~\bibnamefont
  {Maga\~na}}, \bibinfo {author} {\bibfnamefont {A.}~\bibnamefont
  {Hernandez-Almada}}, \ and\ \bibinfo {author} {\bibfnamefont
  {V.}~\bibnamefont {Motta}},\ }\href {\doibase 10.1142/S0218271818500062}
  {\bibfield  {journal} {\bibinfo  {journal} {IJMPD}\ }\textbf {\bibinfo
  {volume} {27}},\ \bibinfo {pages} {18560006} (\bibinfo {year} {2018})},\
  \Eprint {http://arxiv.org/abs/1609.08220} {arXiv:1609.08220 [astro-ph.CO]}
  \BibitemShut {NoStop}%
\bibitem [{\citenamefont {Hoff~da Silva}(2011)}]{HoffdaSilva:2011bd}%
  \BibitemOpen
  \bibfield  {author} {\bibinfo {author} {\bibfnamefont {J.~M.}\ \bibnamefont
  {Hoff~da Silva}},\ }\href {\doibase 10.1103/PhysRevD.83.066001} {\bibfield
  {journal} {\bibinfo  {journal} {Phys. Rev.}\ }\textbf {\bibinfo {volume}
  {D83}},\ \bibinfo {pages} {066001} (\bibinfo {year} {2011})},\ \Eprint
  {http://arxiv.org/abs/1101.4214} {arXiv:1101.4214 [gr-qc]} \BibitemShut
  {NoStop}%
\bibitem [{\citenamefont {Guendelman}\ \emph {et~al.}()\citenamefont
  {Guendelman}, \citenamefont {Kaganovich}, \citenamefont {Nissimov},\ and\
  \citenamefont {Pacheva}}]{Guendelman:2002mf}%
  \BibitemOpen
  \bibfield  {author} {\bibinfo {author} {\bibfnamefont {E.}~\bibnamefont
  {Guendelman}}, \bibinfo {author} {\bibfnamefont {A.}~\bibnamefont
  {Kaganovich}}, \bibinfo {author} {\bibfnamefont {E.}~\bibnamefont
  {Nissimov}}, \ and\ \bibinfo {author} {\bibfnamefont {S.}~\bibnamefont
  {Pacheva}},\ }in\ \href@noop {} {\emph {\bibinfo {booktitle} {{Proceedings,
  1st Adv. Res. Workshop on Grav., Astro.}}}}\BibitemShut {Stop}%
\bibitem [{\citenamefont {Gergely}(2009)}]{Gergely:2008jr}%
  \BibitemOpen
  \bibfield  {author} {\bibinfo {author} {\bibfnamefont {L.~A.}\ \bibnamefont
  {Gergely}},\ }\href {\doibase 10.1103/PhysRevD.79.086007} {\bibfield
  {journal} {\bibinfo  {journal} {Phys. Rev.}\ }\textbf {\bibinfo {volume}
  {D79}},\ \bibinfo {pages} {086007} (\bibinfo {year} {2009})},\ \Eprint
  {http://arxiv.org/abs/0806.4006} {arXiv:0806.4006 [gr-qc]} \BibitemShut
  {NoStop}%
\bibitem [{\citenamefont {Aros}\ and\ \citenamefont
  {Estrada}(2017)}]{Aros:2016wpv}%
  \BibitemOpen
  \bibfield  {author} {\bibinfo {author} {\bibfnamefont {R.}~\bibnamefont
  {Aros}}\ and\ \bibinfo {author} {\bibfnamefont {M.}~\bibnamefont {Estrada}},\
  }\href {\doibase 10.1088/0253-6102/68/5/595} {\bibfield  {journal} {\bibinfo
  {journal} {Commun. Theor. Phys.}\ }\textbf {\bibinfo {volume} {68}},\
  \bibinfo {pages} {595} (\bibinfo {year} {2017})},\ \Eprint
  {http://arxiv.org/abs/1603.05337} {arXiv:1603.05337 [gr-qc]} \BibitemShut
  {NoStop}%
\bibitem [{\citenamefont {Bazeia}\ \emph {et~al.}(2014)\citenamefont {Bazeia},
  \citenamefont {Hoff~da Silva},\ and\ \citenamefont
  {da~Rocha}}]{Bazeia:2014tua}%
  \BibitemOpen
  \bibfield  {author} {\bibinfo {author} {\bibfnamefont {D.}~\bibnamefont
  {Bazeia}}, \bibinfo {author} {\bibfnamefont {J.~M.}\ \bibnamefont {Hoff~da
  Silva}}, \ and\ \bibinfo {author} {\bibfnamefont {R.}~\bibnamefont
  {da~Rocha}},\ }\href {\doibase 10.1103/PhysRevD.90.047902} {\bibfield
  {journal} {\bibinfo  {journal} {Phys. Rev.}\ }\textbf {\bibinfo {volume}
  {D90}},\ \bibinfo {pages} {047902} (\bibinfo {year} {2014})},\ \Eprint
  {http://arxiv.org/abs/1401.6985} {arXiv:1401.6985 [hep-th]} \BibitemShut
  {NoStop}%
\bibitem [{\citenamefont {Casadio}\ \emph {et~al.}(2014)\citenamefont
  {Casadio}, \citenamefont {Ovalle},\ and\ \citenamefont
  {da~Rocha}}]{Casadio:2013uma}%
  \BibitemOpen
  \bibfield  {author} {\bibinfo {author} {\bibfnamefont {R.}~\bibnamefont
  {Casadio}}, \bibinfo {author} {\bibfnamefont {J.}~\bibnamefont {Ovalle}}, \
  and\ \bibinfo {author} {\bibfnamefont {R.}~\bibnamefont {da~Rocha}},\ }\href
  {\doibase 10.1088/0264-9381/31/4/045016} {\bibfield  {journal} {\bibinfo
  {journal} {Class. Quant. Grav.}\ }\textbf {\bibinfo {volume} {31}},\ \bibinfo
  {pages} {045016} (\bibinfo {year} {2014})},\ \Eprint
  {http://arxiv.org/abs/1310.5853} {arXiv:1310.5853 [gr-qc]} \BibitemShut
  {NoStop}%
\bibitem [{\citenamefont {da~Rocha}\ and\ \citenamefont {Hoff~da
  Silva}(2012)}]{daRocha:2012pt}%
  \BibitemOpen
  \bibfield  {author} {\bibinfo {author} {\bibfnamefont {R.}~\bibnamefont
  {da~Rocha}}\ and\ \bibinfo {author} {\bibfnamefont {J.~M.}\ \bibnamefont
  {Hoff~da Silva}},\ }\href {\doibase 10.1103/PhysRevD.85.046009} {\bibfield
  {journal} {\bibinfo  {journal} {Phys. Rev.}\ }\textbf {\bibinfo {volume}
  {D85}},\ \bibinfo {pages} {046009} (\bibinfo {year} {2012})},\ \Eprint
  {http://arxiv.org/abs/1202.1256} {arXiv:1202.1256 [gr-qc]} \BibitemShut
  {NoStop}%
\bibitem [{\citenamefont {Abdalla}\ \emph {et~al.}(2009)\citenamefont
  {Abdalla}, \citenamefont {da~Silva},\ and\ \citenamefont
  {da~Rocha}}]{PhysRevD.80.046003}%
  \BibitemOpen
  \bibfield  {author} {\bibinfo {author} {\bibfnamefont {M.~C.~B.}\
  \bibnamefont {Abdalla}}, \bibinfo {author} {\bibfnamefont {J.~M.~H.}\
  \bibnamefont {da~Silva}}, \ and\ \bibinfo {author} {\bibfnamefont
  {R.}~\bibnamefont {da~Rocha}},\ }\href {\doibase 10.1103/PhysRevD.80.046003}
  {\bibfield  {journal} {\bibinfo  {journal} {Phys. Rev. D}\ }\textbf {\bibinfo
  {volume} {80}},\ \bibinfo {pages} {046003} (\bibinfo {year}
  {2009})}\BibitemShut {NoStop}%
\bibitem [{\citenamefont {{Casadio}}\ and\ \citenamefont {{da
  Rocha}}(2016)}]{Casadio:2016P}%
  \BibitemOpen
  \bibfield  {author} {\bibinfo {author} {\bibfnamefont {R.}~\bibnamefont
  {{Casadio}}}\ and\ \bibinfo {author} {\bibfnamefont {R.}~\bibnamefont {{da
  Rocha}}},\ }\href {\doibase 10.1016/j.physletb.2016.10.072} {\bibfield
  {journal} {\bibinfo  {journal} {Physics Letters B}\ }\textbf {\bibinfo
  {volume} {763}},\ \bibinfo {pages} {434} (\bibinfo {year} {2016})},\ \Eprint
  {http://arxiv.org/abs/1610.01572} {arXiv:1610.01572 [hep-th]} \BibitemShut
  {NoStop}%
\bibitem [{\citenamefont {Gergely}(2008)}]{Gergely:2008fw}%
  \BibitemOpen
  \bibfield  {author} {\bibinfo {author} {\bibfnamefont {L.~A.}\ \bibnamefont
  {Gergely}},\ }\href {\doibase 10.1103/PhysRevD.78.084006} {\bibfield
  {journal} {\bibinfo  {journal} {Phys. Rev.}\ }\textbf {\bibinfo {volume}
  {D78}},\ \bibinfo {pages} {084006} (\bibinfo {year} {2008})},\ \Eprint
  {http://arxiv.org/abs/0806.3857} {arXiv:0806.3857 [gr-qc]} \BibitemShut
  {NoStop}%
\bibitem [{\citenamefont {Wong}\ \emph {et~al.}(2010)\citenamefont {Wong},
  \citenamefont {Cheng},\ and\ \citenamefont {Harko}}]{Wong:2010rg}%
  \BibitemOpen
  \bibfield  {author} {\bibinfo {author} {\bibfnamefont {K.~C.}\ \bibnamefont
  {Wong}}, \bibinfo {author} {\bibfnamefont {K.~S.}\ \bibnamefont {Cheng}}, \
  and\ \bibinfo {author} {\bibfnamefont {T.}~\bibnamefont {Harko}},\ }\href
  {\doibase 10.1140/epjc/s10052-010-1348-9} {\bibfield  {journal} {\bibinfo
  {journal} {Eur. Phys. J.}\ }\textbf {\bibinfo {volume} {C68}},\ \bibinfo
  {pages} {241} (\bibinfo {year} {2010})},\ \Eprint
  {http://arxiv.org/abs/1005.3101} {arXiv:1005.3101 [gr-qc]} \BibitemShut
  {NoStop}%
\bibitem [{\citenamefont {Shiromizu}\ \emph {et~al.}(2000)\citenamefont
  {Shiromizu}, \citenamefont {Maeda},\ and\ \citenamefont {Sasaki}}]{sms}%
  \BibitemOpen
  \bibfield  {author} {\bibinfo {author} {\bibfnamefont {T.}~\bibnamefont
  {Shiromizu}}, \bibinfo {author} {\bibfnamefont {K.}~\bibnamefont {Maeda}}, \
  and\ \bibinfo {author} {\bibfnamefont {M.}~\bibnamefont {Sasaki}},\ }\href
  {http://link.aps.org/doi/10.1103/PhysRevD.62.024012} {\bibfield  {journal}
  {\bibinfo  {journal} {Phys. Rev. D}\ }\textbf {\bibinfo {volume} {62}},\
  \bibinfo {pages} {024012} (\bibinfo {year} {2000})}\BibitemShut {NoStop}%
\bibitem [{\citenamefont {Maartens}(2000)}]{m2000}%
  \BibitemOpen
  \bibfield  {author} {\bibinfo {author} {\bibfnamefont {R.}~\bibnamefont
  {Maartens}},\ }\href@noop {} {\bibfield  {journal} {\bibinfo  {journal}
  {Phys. Rev. D}\ }\textbf {\bibinfo {volume} {62}},\ \bibinfo {pages} {084023}
  (\bibinfo {year} {2000})}\BibitemShut {NoStop}%
\bibitem [{\citenamefont {{Foreman-Mackey}}\ \emph {et~al.}(2013)\citenamefont
  {{Foreman-Mackey}}, \citenamefont {{Hogg}}, \citenamefont {{Lang}},\ and\
  \citenamefont {{Goodman}}}]{Foreman:2013}%
  \BibitemOpen
  \bibfield  {author} {\bibinfo {author} {\bibfnamefont {D.}~\bibnamefont
  {{Foreman-Mackey}}}, \bibinfo {author} {\bibfnamefont {D.~W.}\ \bibnamefont
  {{Hogg}}}, \bibinfo {author} {\bibfnamefont {D.}~\bibnamefont {{Lang}}}, \
  and\ \bibinfo {author} {\bibfnamefont {J.}~\bibnamefont {{Goodman}}},\ }\href
  {\doibase 10.1086/670067} {\bibfield  {journal} {\bibinfo  {journal} {PASP}\
  }\textbf {\bibinfo {volume} {125}},\ \bibinfo {pages} {306} (\bibinfo {year}
  {2013})},\ \Eprint {http://arxiv.org/abs/1202.3665} {arXiv:1202.3665
  [astro-ph.IM]} \BibitemShut {NoStop}%
\bibitem [{\citenamefont {Riess}\ \emph {et~al.}(2016)\citenamefont {Riess}
  \emph {et~al.}}]{Riess:2016jrr}%
  \BibitemOpen
  \bibfield  {author} {\bibinfo {author} {\bibfnamefont {A.~G.}\ \bibnamefont
  {Riess}} \emph {et~al.},\ }\href {\doibase 10.3847/0004-637X/826/1/56}
  {\bibfield  {journal} {\bibinfo  {journal} {Astrophys. J.}\ }\textbf
  {\bibinfo {volume} {826}},\ \bibinfo {pages} {56} (\bibinfo {year} {2016})},\
  \Eprint {http://arxiv.org/abs/1604.01424} {arXiv:1604.01424 [astro-ph.CO]}
  \BibitemShut {NoStop}%
\bibitem [{\citenamefont {{Wei}}\ \emph {et~al.}(2017)\citenamefont {{Wei}},
  \citenamefont {{Melia}},\ and\ \citenamefont {{Wu}}}]{Wei:2017}%
  \BibitemOpen
  \bibfield  {author} {\bibinfo {author} {\bibfnamefont {J.-J.}\ \bibnamefont
  {{Wei}}}, \bibinfo {author} {\bibfnamefont {F.}~\bibnamefont {{Melia}}}, \
  and\ \bibinfo {author} {\bibfnamefont {X.-F.}\ \bibnamefont {{Wu}}},\ }\href
  {\doibase 10.3847/1538-4357/835/2/270} {\bibfield  {journal} {\bibinfo
  {journal} {\apj}\ }\textbf {\bibinfo {volume} {835}},\ \bibinfo {eid} {270}
  (\bibinfo {year} {2017})},\ \Eprint {http://arxiv.org/abs/1612.08491}
  {arXiv:1612.08491} \BibitemShut {NoStop}%
\bibitem [{\citenamefont {{Bernal}}\ \emph {et~al.}(2016)\citenamefont
  {{Bernal}}, \citenamefont {{Verde}},\ and\ \citenamefont
  {{Riess}}}]{Bernal:2016}%
  \BibitemOpen
  \bibfield  {author} {\bibinfo {author} {\bibfnamefont {J.~L.}\ \bibnamefont
  {{Bernal}}}, \bibinfo {author} {\bibfnamefont {L.}~\bibnamefont {{Verde}}}, \
  and\ \bibinfo {author} {\bibfnamefont {A.~G.}\ \bibnamefont {{Riess}}},\
  }\href {\doibase 10.1088/1475-7516/2016/10/019} {\bibfield  {journal}
  {\bibinfo  {journal} {J. Cosmology Astropart. Phys.}\ }\textbf {\bibinfo
  {volume} {10}},\ \bibinfo {eid} {019} (\bibinfo {year} {2016})},\ \Eprint
  {http://arxiv.org/abs/1607.05617} {arXiv:1607.05617} \BibitemShut {NoStop}%
\bibitem [{\citenamefont {Betoule}\ \emph {et~al.}(2014)\citenamefont {Betoule}
  \emph {et~al.}}]{Betoule:2014}%
  \BibitemOpen
  \bibfield  {author} {\bibinfo {author} {\bibfnamefont {M.}~\bibnamefont
  {Betoule}} \emph {et~al.} (\bibinfo {collaboration} {SDSS}),\ }\href
  {\doibase 10.1051/0004-6361/201423413} {\bibfield  {journal} {\bibinfo
  {journal} {Astron. Astrophys.}\ }\textbf {\bibinfo {volume} {568}},\ \bibinfo
  {pages} {A22} (\bibinfo {year} {2014})},\ \Eprint
  {http://arxiv.org/abs/1401.4064} {arXiv:1401.4064 [astro-ph.CO]} \BibitemShut
  {NoStop}%
\bibitem [{\citenamefont {Conley}\ \emph {et~al.}(2007)\citenamefont {Conley},
  \citenamefont {Carlberg}, \citenamefont {Guy}, \citenamefont {Howell},
  \citenamefont {Jha}, \citenamefont {Riess},\ and\ \citenamefont
  {Sullivan}}]{Conley:2007ng}%
  \BibitemOpen
  \bibfield  {author} {\bibinfo {author} {\bibfnamefont {A.~J.}\ \bibnamefont
  {Conley}}, \bibinfo {author} {\bibfnamefont {R.~G.}\ \bibnamefont
  {Carlberg}}, \bibinfo {author} {\bibfnamefont {J.}~\bibnamefont {Guy}},
  \bibinfo {author} {\bibfnamefont {D.~A.}\ \bibnamefont {Howell}}, \bibinfo
  {author} {\bibfnamefont {S.}~\bibnamefont {Jha}}, \bibinfo {author}
  {\bibfnamefont {A.~G.}\ \bibnamefont {Riess}}, \ and\ \bibinfo {author}
  {\bibfnamefont {M.}~\bibnamefont {Sullivan}} (\bibinfo {collaboration}
  {SNLS}),\ }\href {\doibase 10.1086/520625} {\bibfield  {journal} {\bibinfo
  {journal} {Astrophys. J.}\ }\textbf {\bibinfo {volume} {664}},\ \bibinfo
  {pages} {L13} (\bibinfo {year} {2007})},\ \Eprint
  {http://arxiv.org/abs/0705.0367} {arXiv:0705.0367 [astro-ph]} \BibitemShut
  {NoStop}%
\bibitem [{\citenamefont {{Conley}}\ \emph {et~al.}(2011)\citenamefont
  {{Conley}}, \citenamefont {{Guy}}, \citenamefont {{Sullivan}}, \citenamefont
  {{Regnault}}, \citenamefont {{Astier}}, \citenamefont {{Balland}},
  \citenamefont {{Basa}}, \citenamefont {{Carlberg}}, \citenamefont
  {{Fouchez}}, \citenamefont {{Hardin}}, \citenamefont {{Hook}}, \citenamefont
  {{Howell}}, \citenamefont {{Pain}}, \citenamefont {{Palanque-Delabrouille}},
  \citenamefont {{Perrett}}, \citenamefont {{Pritchet}}, \citenamefont
  {{Rich}}, \citenamefont {{Ruhlmann-Kleider}}, \citenamefont {{Balam}},
  \citenamefont {{Baumont}}, \citenamefont {{Ellis}}, \citenamefont {{Fabbro}},
  \citenamefont {{Fakhouri}}, \citenamefont {{Fourmanoit}}, \citenamefont
  {{Gonz{\'a}lez-Gait{\'a}n}}, \citenamefont {{Graham}}, \citenamefont
  {{Hudson}}, \citenamefont {{Hsiao}}, \citenamefont {{Kronborg}},
  \citenamefont {{Lidman}}, \citenamefont {{Mourao}}, \citenamefont {{Neill}},
  \citenamefont {{Perlmutter}}, \citenamefont {{Ripoche}}, \citenamefont
  {{Suzuki}},\ and\ \citenamefont {{Walker}}}]{Conley:2011}%
  \BibitemOpen
  \bibfield  {author} {\bibinfo {author} {\bibfnamefont {A.}~\bibnamefont
  {{Conley}}}, \bibinfo {author} {\bibfnamefont {J.}~\bibnamefont {{Guy}}},
  \bibinfo {author} {\bibfnamefont {M.}~\bibnamefont {{Sullivan}}}, \bibinfo
  {author} {\bibfnamefont {N.}~\bibnamefont {{Regnault}}}, \bibinfo {author}
  {\bibfnamefont {P.}~\bibnamefont {{Astier}}}, \bibinfo {author}
  {\bibfnamefont {C.}~\bibnamefont {{Balland}}}, \bibinfo {author}
  {\bibfnamefont {S.}~\bibnamefont {{Basa}}}, \bibinfo {author} {\bibfnamefont
  {R.~G.}\ \bibnamefont {{Carlberg}}}, \bibinfo {author} {\bibfnamefont
  {D.}~\bibnamefont {{Fouchez}}}, \bibinfo {author} {\bibfnamefont
  {D.}~\bibnamefont {{Hardin}}}, \bibinfo {author} {\bibfnamefont {I.~M.}\
  \bibnamefont {{Hook}}}, \bibinfo {author} {\bibfnamefont {D.~A.}\
  \bibnamefont {{Howell}}}, \bibinfo {author} {\bibfnamefont {R.}~\bibnamefont
  {{Pain}}}, \bibinfo {author} {\bibfnamefont {N.}~\bibnamefont
  {{Palanque-Delabrouille}}}, \bibinfo {author} {\bibfnamefont {K.~M.}\
  \bibnamefont {{Perrett}}}, \bibinfo {author} {\bibfnamefont {C.~J.}\
  \bibnamefont {{Pritchet}}}, \bibinfo {author} {\bibfnamefont
  {J.}~\bibnamefont {{Rich}}}, \bibinfo {author} {\bibfnamefont
  {V.}~\bibnamefont {{Ruhlmann-Kleider}}}, \bibinfo {author} {\bibfnamefont
  {D.}~\bibnamefont {{Balam}}}, \bibinfo {author} {\bibfnamefont
  {S.}~\bibnamefont {{Baumont}}}, \bibinfo {author} {\bibfnamefont {R.~S.}\
  \bibnamefont {{Ellis}}}, \bibinfo {author} {\bibfnamefont {S.}~\bibnamefont
  {{Fabbro}}}, \bibinfo {author} {\bibfnamefont {H.~K.}\ \bibnamefont
  {{Fakhouri}}}, \bibinfo {author} {\bibfnamefont {N.}~\bibnamefont
  {{Fourmanoit}}}, \bibinfo {author} {\bibfnamefont {S.}~\bibnamefont
  {{Gonz{\'a}lez-Gait{\'a}n}}}, \bibinfo {author} {\bibfnamefont {M.~L.}\
  \bibnamefont {{Graham}}}, \bibinfo {author} {\bibfnamefont {M.~J.}\
  \bibnamefont {{Hudson}}}, \bibinfo {author} {\bibfnamefont {E.}~\bibnamefont
  {{Hsiao}}}, \bibinfo {author} {\bibfnamefont {T.}~\bibnamefont {{Kronborg}}},
  \bibinfo {author} {\bibfnamefont {C.}~\bibnamefont {{Lidman}}}, \bibinfo
  {author} {\bibfnamefont {A.~M.}\ \bibnamefont {{Mourao}}}, \bibinfo {author}
  {\bibfnamefont {J.~D.}\ \bibnamefont {{Neill}}}, \bibinfo {author}
  {\bibfnamefont {S.}~\bibnamefont {{Perlmutter}}}, \bibinfo {author}
  {\bibfnamefont {P.}~\bibnamefont {{Ripoche}}}, \bibinfo {author}
  {\bibfnamefont {N.}~\bibnamefont {{Suzuki}}}, \ and\ \bibinfo {author}
  {\bibfnamefont {E.~S.}\ \bibnamefont {{Walker}}},\ }\href {\doibase
  10.1088/0067-0049/192/1/1} {\bibfield  {journal} {\bibinfo  {journal}
  {Astrophys. J. Supp.}\ }\textbf {\bibinfo {volume} {192}},\ \bibinfo {eid}
  {1} (\bibinfo {year} {2011})},\ \Eprint {http://arxiv.org/abs/1104.1443}
  {arXiv:1104.1443 [astro-ph.CO]} \BibitemShut {NoStop}%
\bibitem [{\citenamefont {Scolnic}\ \emph
  {et~al.}(2014{\natexlab{a}})\citenamefont {Scolnic} \emph
  {et~al.}}]{Scolnic:2013efb}%
  \BibitemOpen
  \bibfield  {author} {\bibinfo {author} {\bibfnamefont {D.}~\bibnamefont
  {Scolnic}} \emph {et~al.},\ }\href {\doibase 10.1088/0004-637X/795/1/45}
  {\bibfield  {journal} {\bibinfo  {journal} {Astrophys. J.}\ }\textbf
  {\bibinfo {volume} {795}},\ \bibinfo {pages} {45} (\bibinfo {year}
  {2014}{\natexlab{a}})},\ \Eprint {http://arxiv.org/abs/1310.3824}
  {arXiv:1310.3824 [astro-ph.CO]} \BibitemShut {NoStop}%
\bibitem [{\citenamefont {Scolnic}\ \emph
  {et~al.}(2014{\natexlab{b}})\citenamefont {Scolnic}, \citenamefont {Riess},
  \citenamefont {Foley}, \citenamefont {Rest}, \citenamefont {Rodney},
  \citenamefont {Brout},\ and\ \citenamefont {Jones}}]{Scolnic:2013xra}%
  \BibitemOpen
  \bibfield  {author} {\bibinfo {author} {\bibfnamefont {D.~M.}\ \bibnamefont
  {Scolnic}}, \bibinfo {author} {\bibfnamefont {A.~G.}\ \bibnamefont {Riess}},
  \bibinfo {author} {\bibfnamefont {R.~J.}\ \bibnamefont {Foley}}, \bibinfo
  {author} {\bibfnamefont {A.}~\bibnamefont {Rest}}, \bibinfo {author}
  {\bibfnamefont {S.~A.}\ \bibnamefont {Rodney}}, \bibinfo {author}
  {\bibfnamefont {D.~J.}\ \bibnamefont {Brout}}, \ and\ \bibinfo {author}
  {\bibfnamefont {D.~O.}\ \bibnamefont {Jones}},\ }\href {\doibase
  10.1088/0004-637X/780/1/37} {\bibfield  {journal} {\bibinfo  {journal}
  {Astrophys. J.}\ }\textbf {\bibinfo {volume} {780}},\ \bibinfo {pages} {37}
  (\bibinfo {year} {2014}{\natexlab{b}})},\ \Eprint
  {http://arxiv.org/abs/1306.4050} {arXiv:1306.4050 [astro-ph.CO]} \BibitemShut
  {NoStop}%
\bibitem [{\citenamefont {Mosher}\ \emph {et~al.}(2014)\citenamefont {Mosher}
  \emph {et~al.}}]{Mosher:2014}%
  \BibitemOpen
  \bibfield  {author} {\bibinfo {author} {\bibfnamefont {J.}~\bibnamefont
  {Mosher}} \emph {et~al.},\ }\href {\doibase 10.1088/0004-637X/793/1/16}
  {\bibfield  {journal} {\bibinfo  {journal} {Astrophys. J.}\ }\textbf
  {\bibinfo {volume} {793}},\ \bibinfo {pages} {16} (\bibinfo {year} {2014})},\
  \Eprint {http://arxiv.org/abs/1401.4065} {arXiv:1401.4065 [astro-ph.CO]}
  \BibitemShut {NoStop}%
\bibitem [{\citenamefont {Beutler}\ \emph {et~al.}(2011)\citenamefont
  {Beutler}, \citenamefont {Blake}, \citenamefont {Colless}, \citenamefont
  {Jones}, \citenamefont {Staveley-Smith}, \citenamefont {Campbell},
  \citenamefont {Parker}, \citenamefont {Saunders},\ and\ \citenamefont
  {Watson}}]{Beutler2011:6dF}%
  \BibitemOpen
  \bibfield  {author} {\bibinfo {author} {\bibfnamefont {F.}~\bibnamefont
  {Beutler}}, \bibinfo {author} {\bibfnamefont {C.}~\bibnamefont {Blake}},
  \bibinfo {author} {\bibfnamefont {M.}~\bibnamefont {Colless}}, \bibinfo
  {author} {\bibfnamefont {D.~H.}\ \bibnamefont {Jones}}, \bibinfo {author}
  {\bibfnamefont {L.}~\bibnamefont {Staveley-Smith}}, \bibinfo {author}
  {\bibfnamefont {L.}~\bibnamefont {Campbell}}, \bibinfo {author}
  {\bibfnamefont {Q.}~\bibnamefont {Parker}}, \bibinfo {author} {\bibfnamefont
  {W.}~\bibnamefont {Saunders}}, \ and\ \bibinfo {author} {\bibfnamefont
  {F.}~\bibnamefont {Watson}},\ }\href {\doibase
  10.1111/j.1365-2966.2011.19250.x} {\bibfield  {journal} {\bibinfo  {journal}
  {Mon. Not. Roy. Astron. Soc.}\ }\textbf {\bibinfo {volume} {416}},\ \bibinfo
  {pages} {3017} (\bibinfo {year} {2011})},\ \Eprint
  {http://arxiv.org/abs/1106.3366} {arXiv:1106.3366 [astro-ph.CO]} \BibitemShut
  {NoStop}%
\bibitem [{\citenamefont {Kazin}\ \emph {et~al.}(2014)\citenamefont {Kazin}
  \emph {et~al.}}]{Kazin:2014}%
  \BibitemOpen
  \bibfield  {author} {\bibinfo {author} {\bibfnamefont {E.~A.}\ \bibnamefont
  {Kazin}} \emph {et~al.},\ }\href {\doibase 10.1093/mnras/stu778} {\bibfield
  {journal} {\bibinfo  {journal} {Mon. Not. Roy. Astron. Soc.}\ }\textbf
  {\bibinfo {volume} {441}},\ \bibinfo {pages} {3524} (\bibinfo {year}
  {2014})},\ \Eprint {http://arxiv.org/abs/1401.0358} {arXiv:1401.0358
  [astro-ph.CO]} \BibitemShut {NoStop}%
\bibitem [{\citenamefont {Gong}\ \emph {et~al.}(2015)\citenamefont {Gong},
  \citenamefont {Ma}, \citenamefont {Zhang},\ and\ \citenamefont
  {Chen}}]{Gong:2015}%
  \BibitemOpen
  \bibfield  {author} {\bibinfo {author} {\bibfnamefont {Y.}~\bibnamefont
  {Gong}}, \bibinfo {author} {\bibfnamefont {Y.-Z.}\ \bibnamefont {Ma}},
  \bibinfo {author} {\bibfnamefont {S.-N.}\ \bibnamefont {Zhang}}, \ and\
  \bibinfo {author} {\bibfnamefont {X.}~\bibnamefont {Chen}},\ }\href {\doibase
  10.1103/PhysRevD.92.109905, 10.1103/PhysRevD.92.063523} {\bibfield  {journal}
  {\bibinfo  {journal} {Phys. Rev.}\ }\textbf {\bibinfo {volume} {D92}},\
  \bibinfo {pages} {063523} (\bibinfo {year} {2015})},\ \bibinfo {note}
  {[Addendum: Phys. Rev.D92,no.10,109905(2015)]},\ \Eprint
  {http://arxiv.org/abs/1505.03584} {arXiv:1505.03584 [astro-ph.CO]}
  \BibitemShut {NoStop}%
\bibitem [{\citenamefont {Ross}\ \emph {et~al.}(2015)\citenamefont {Ross},
  \citenamefont {Samushia}, \citenamefont {Howlett}, \citenamefont {Percival},
  \citenamefont {Burden},\ and\ \citenamefont {Manera}}]{Ross:2014}%
  \BibitemOpen
  \bibfield  {author} {\bibinfo {author} {\bibfnamefont {A.~J.}\ \bibnamefont
  {Ross}}, \bibinfo {author} {\bibfnamefont {L.}~\bibnamefont {Samushia}},
  \bibinfo {author} {\bibfnamefont {C.}~\bibnamefont {Howlett}}, \bibinfo
  {author} {\bibfnamefont {W.~J.}\ \bibnamefont {Percival}}, \bibinfo {author}
  {\bibfnamefont {A.}~\bibnamefont {Burden}}, \ and\ \bibinfo {author}
  {\bibfnamefont {M.}~\bibnamefont {Manera}},\ }\href {\doibase
  10.1093/mnras/stv154} {\bibfield  {journal} {\bibinfo  {journal} {Mon. Not.
  Roy. Astron. Soc.}\ }\textbf {\bibinfo {volume} {449}},\ \bibinfo {pages}
  {835} (\bibinfo {year} {2015})},\ \Eprint {http://arxiv.org/abs/1409.3242}
  {arXiv:1409.3242 [astro-ph.CO]} \BibitemShut {NoStop}%
\bibitem [{\citenamefont {Anderson}\ \emph {et~al.}(2014)\citenamefont
  {Anderson} \emph {et~al.}}]{Anderson:2014}%
  \BibitemOpen
  \bibfield  {author} {\bibinfo {author} {\bibfnamefont {L.}~\bibnamefont
  {Anderson}} \emph {et~al.} (\bibinfo {collaboration} {BOSS}),\ }\href
  {\doibase 10.1093/mnras/stu523} {\bibfield  {journal} {\bibinfo  {journal}
  {Mon. Not. Roy. Astron. Soc.}\ }\textbf {\bibinfo {volume} {441}},\ \bibinfo
  {pages} {24} (\bibinfo {year} {2014})},\ \Eprint
  {http://arxiv.org/abs/1312.4877} {arXiv:1312.4877 [astro-ph.CO]} \BibitemShut
  {NoStop}%
\bibitem [{\citenamefont {Bautista}\ \emph {et~al.}(2017)\citenamefont
  {Bautista} \emph {et~al.}}]{Bautista:2017}%
  \BibitemOpen
  \bibfield  {author} {\bibinfo {author} {\bibfnamefont {J.~E.}\ \bibnamefont
  {Bautista}} \emph {et~al.},\ }\href {\doibase 10.1051/0004-6361/201730533}
  {\bibfield  {journal} {\bibinfo  {journal} {Astron. Astrophys.}\ }\textbf
  {\bibinfo {volume} {603}},\ \bibinfo {pages} {A12} (\bibinfo {year}
  {2017})},\ \Eprint {http://arxiv.org/abs/1702.00176} {arXiv:1702.00176
  [astro-ph.CO]} \BibitemShut {NoStop}%
\bibitem [{\citenamefont {Neveu}\ \emph {et~al.}(2016)\citenamefont {Neveu},
  \citenamefont {Ruhlmann-Kleider}, \citenamefont {Astier}, \citenamefont
  {Besanon}, \citenamefont {Guy}, \citenamefont {Muller},\ and\ \citenamefont
  {Babichev}}]{Neveu:2016}%
  \BibitemOpen
  \bibfield  {author} {\bibinfo {author} {\bibfnamefont {J.}~\bibnamefont
  {Neveu}}, \bibinfo {author} {\bibfnamefont {V.}~\bibnamefont
  {Ruhlmann-Kleider}}, \bibinfo {author} {\bibfnamefont {P.}~\bibnamefont
  {Astier}}, \bibinfo {author} {\bibfnamefont {M.}~\bibnamefont {Besanon}},
  \bibinfo {author} {\bibfnamefont {J.}~\bibnamefont {Guy}}, \bibinfo {author}
  {\bibfnamefont {A.}~\bibnamefont {Muller}}, \ and\ \bibinfo {author}
  {\bibfnamefont {E.}~\bibnamefont {Babichev}},\ }\href@noop {} {\  (\bibinfo
  {year} {2016})},\ \Eprint {http://arxiv.org/abs/1605.02627} {arXiv:1605.02627
  [gr-qc]} \BibitemShut {NoStop}%
\bibitem [{\citenamefont {Alam}\ \emph
  {et~al.}(2017{\natexlab{b}})\citenamefont {Alam}, \citenamefont {Bag},\ and\
  \citenamefont {Sahni}}]{PhysRevD.95.023524}%
  \BibitemOpen
  \bibfield  {author} {\bibinfo {author} {\bibfnamefont {U.}~\bibnamefont
  {Alam}}, \bibinfo {author} {\bibfnamefont {S.}~\bibnamefont {Bag}}, \ and\
  \bibinfo {author} {\bibfnamefont {V.}~\bibnamefont {Sahni}},\ }\href
  {\doibase 10.1103/PhysRevD.95.023524} {\bibfield  {journal} {\bibinfo
  {journal} {Phys. Rev. D}\ }\textbf {\bibinfo {volume} {95}},\ \bibinfo
  {pages} {023524} (\bibinfo {year} {2017}{\natexlab{b}})}\BibitemShut
  {NoStop}%
\bibitem [{\citenamefont {Germani}\ and\ \citenamefont
  {Maartens}(2001)}]{Germani:2001du}%
  \BibitemOpen
  \bibfield  {author} {\bibinfo {author} {\bibfnamefont {C.}~\bibnamefont
  {Germani}}\ and\ \bibinfo {author} {\bibfnamefont {R.}~\bibnamefont
  {Maartens}},\ }\href {\doibase 10.1103/PhysRevD.64.124010} {\bibfield
  {journal} {\bibinfo  {journal} {Phys. Rev.}\ }\textbf {\bibinfo {volume}
  {D64}},\ \bibinfo {pages} {124010} (\bibinfo {year} {2001})},\ \Eprint
  {http://arxiv.org/abs/hep-th/0107011} {arXiv:hep-th/0107011 [hep-th]}
  \BibitemShut {NoStop}%
\bibitem [{\citenamefont {Maartens}(2004)}]{Maartens:2003tw}%
  \BibitemOpen
  \bibfield  {author} {\bibinfo {author} {\bibfnamefont {R.}~\bibnamefont
  {Maartens}},\ }\href@noop {} {\bibfield  {journal} {\bibinfo  {journal}
  {Living Rev. Rel.}\ }\textbf {\bibinfo {volume} {7}},\ \bibinfo {pages} {7}
  (\bibinfo {year} {2004})},\ \Eprint {http://arxiv.org/abs/gr-qc/0312059}
  {arXiv:gr-qc/0312059 [gr-qc]} \BibitemShut {NoStop}%
\bibitem [{\citenamefont {{Akaike}}(1974)}]{Akaike:1974}%
  \BibitemOpen
  \bibfield  {author} {\bibinfo {author} {\bibfnamefont {H.}~\bibnamefont
  {{Akaike}}},\ }\href@noop {} {\bibfield  {journal} {\bibinfo  {journal} {IEEE
  Transactions on Automatic Control}\ }\textbf {\bibinfo {volume} {19}},\
  \bibinfo {pages} {716} (\bibinfo {year} {1974})}\BibitemShut {NoStop}%
\bibitem [{\citenamefont {Shi}\ \emph {et~al.}(2012)\citenamefont {Shi},
  \citenamefont {Huang},\ and\ \citenamefont {Lu}}]{Shi:2012ma}%
  \BibitemOpen
  \bibfield  {author} {\bibinfo {author} {\bibfnamefont {K.}~\bibnamefont
  {Shi}}, \bibinfo {author} {\bibfnamefont {Y.}~\bibnamefont {Huang}}, \ and\
  \bibinfo {author} {\bibfnamefont {T.}~\bibnamefont {Lu}},\ }\href {\doibase
  10.1111/j.1365-2966.2012.21784.x} {\bibfield  {journal} {\bibinfo  {journal}
  {Mon. Not. Roy. Astron. Soc.}\ }\textbf {\bibinfo {volume} {426}},\ \bibinfo
  {pages} {2452} (\bibinfo {year} {2012})},\ \Eprint
  {http://arxiv.org/abs/1207.5875} {arXiv:1207.5875 [astro-ph.CO]} \BibitemShut
  {NoStop}%
\bibitem [{\citenamefont {{Schwarz}}(1978)}]{Schwarz:1978}%
  \BibitemOpen
  \bibfield  {author} {\bibinfo {author} {\bibfnamefont {G.}~\bibnamefont
  {{Schwarz}}},\ }\href@noop {} {\bibfield  {journal} {\bibinfo  {journal}
  {Annals of Statistics}\ }\textbf {\bibinfo {volume} {6}},\ \bibinfo {pages}
  {461} (\bibinfo {year} {1978})}\BibitemShut {NoStop}%
\bibitem [{\citenamefont {Jarosz}\ and\ \citenamefont
  {Wiley}(2014)}]{Jarosz:2014}%
  \BibitemOpen
  \bibfield  {author} {\bibinfo {author} {\bibfnamefont {A.~F.}\ \bibnamefont
  {Jarosz}}\ and\ \bibinfo {author} {\bibfnamefont {J.}~\bibnamefont {Wiley}}
  (\bibinfo {collaboration} {SNLS}),\ }\href {\doibase 10.7771/1932-6246.1167}
  {\bibfield  {journal} {\bibinfo  {journal} {Journal of Problem Solving}\
  }\textbf {\bibinfo {volume} {7}} (\bibinfo {year} {2014}),\
  10.7771/1932-6246.1167}\BibitemShut {NoStop}%
\bibitem [{\citenamefont {Zhao}\ \emph {et~al.}(2017)\citenamefont {Zhao} \emph
  {et~al.}}]{Zhao:2017cud}%
  \BibitemOpen
  \bibfield  {author} {\bibinfo {author} {\bibfnamefont {G.-B.}\ \bibnamefont
  {Zhao}} \emph {et~al.},\ }\href {\doibase 10.1038/s41550-017-0216-z}
  {\bibfield  {journal} {\bibinfo  {journal} {Nat. Astron.}\ }\textbf {\bibinfo
  {volume} {1}},\ \bibinfo {pages} {627} (\bibinfo {year} {2017})},\ \Eprint
  {http://arxiv.org/abs/1701.08165} {arXiv:1701.08165 [astro-ph.CO]}
  \BibitemShut {NoStop}%
\bibitem [{\citenamefont {Koyama}(2003)}]{Koyama:2003be}%
  \BibitemOpen
  \bibfield  {author} {\bibinfo {author} {\bibfnamefont {K.}~\bibnamefont
  {Koyama}},\ }\href {\doibase 10.1103/PhysRevLett.91.221301} {\bibfield
  {journal} {\bibinfo  {journal} {Phys. Rev. Lett.}\ }\textbf {\bibinfo
  {volume} {91}},\ \bibinfo {pages} {221301} (\bibinfo {year} {2003})},\
  \Eprint {http://arxiv.org/abs/astro-ph/0303108} {arXiv:astro-ph/0303108
  [astro-ph]} \BibitemShut {NoStop}%
\end{thebibliography}%

\end{document}